\documentclass[graybox]{svmult}

\usepackage{type1cm}        
\usepackage{makeidx}         
\usepackage{graphicx}        
\usepackage{multicol}        
\usepackage[bottom]{footmisc}

\usepackage{newtxtext}       %
\usepackage[varvw]{newtxmath}       

\usepackage[utf8]{inputenc}
\usepackage{hyperref}
\usepackage{color}
\usepackage{epsfig}
\usepackage{graphics}
\usepackage{epsf}
\usepackage{stmaryrd}
\usepackage{multirow,array}
\usepackage{url}
\usepackage{wasysym}
\usepackage[small]{caption}
\usepackage{enumitem}

\makeindex             

\begin{document}

\title*{Causal Inference in Recommender Systems: A Survey of Strategies for Bias Mitigation, Explanation, and Generalization}
\author{Yaochen Zhu, Jing Ma, and Jundong Li}
\institute{Yaochen Zhu \at Department of Electrical and Computer Engineering, University of Virginia, \email{uqp4qh@virginia.edu}
\and Jing Ma \at Department of Computer Science, University of Virginia, \email{jm3mr@virginia.edu}
\and Jundong Li \at Department of Electrical and Computer Engineering, Department of Computer Science, and School of Data Science, University of Virginia, \email{jl6qk@virginia.edu}}
%
%
\titlerunning{Causal Inference for Recommender Systems: A survey}
\maketitle

\abstract*{In the era of information overload, recommender systems (RSs) have become an indispensable part of online service platforms. Traditional RSs estimate user interests and predict their future behaviors by utilizing correlations in the observational historical activities, their profiles, and the content of interacted items. However, since the inherent causal reasons that lead to the observed users' behaviors are not considered, multiple types of biases could exist in the generated recommendations. In addition, the causal motives that drive user activities are usually entangled in these RSs, where the explainability and generalization abilities of recommendations cannot be guaranteed. To address these drawbacks, recent years have witnessed an upsurge of interest in enhancing traditional RSs with causal inference techniques. In this survey, we provide a systematic overview of causal RSs and help readers gain a comprehensive understanding of this promising area. We start with the basic concepts of traditional RSs and their limitations due to the lack of causal reasoning ability. We then discuss how different causal inference techniques can be introduced to address these challenges, with an emphasis on debiasing, explainability promotion, and generalization improvement. Furthermore, we thoroughly analyze various evaluation strategies for causal RSs, focusing especially on how to reliably estimate their performance with biased data if the causal effects of interests are unavailable. Finally, we provide insights into potential directions for future causal RS research.}

\abstract{In the era of information overload, recommender systems (RSs) have become an indispensable part of online service platforms. Traditional RSs estimate user interests and predict their future behaviors by utilizing correlations in the observational user historical activities, user profiles, and the content of interacted items. However, since the inherent causal reasons that lead to the observed user behaviors are not considered, multiple types of biases could exist in the generated recommendations. In addition, the causal motives that drive user activities are usually entangled in these RSs, where the explainability and generalization abilities of recommendations cannot be guaranteed. To address these drawbacks, recent years have witnessed an upsurge of interest in enhancing traditional RSs with causal inference techniques. In this survey, we provide a systematic overview of causal RSs and help readers gain a comprehensive understanding of this promising area. We start with the basic concepts of traditional RSs and their limitations due to the lack of causal reasoning ability. We then discuss how different causal inference techniques can be introduced to address these challenges, with an emphasis on debiasing, explainability promotion, and generalization improvement. Furthermore, we thoroughly analyze various evaluation strategies for causal RSs, focusing especially on how to reliably estimate their performance with biased data if the causal effects of interests are unavailable. Finally, we provide insights into potential directions for future causal RS research.}

\section{Introduction}

With information growing exponentially on the web, recommender systems (RSs) are playing an increasingly pivotal role in modern online services, due to their ability to automatically deliver items\footnote{We use the term item in a broad sense to refer to anything recommendable to users, such as news \cite{liu2010personalized}, jobs \cite{paparrizos2011machine}, articles \cite{wang2011collaborative}, music \cite{yi2021cross}, movies \cite{harper2015movielens}, micro-videos \cite{xie2020multimodal}, PoIs \cite{ye2011exploiting}, hashtags \cite{gong2016hashtag}, etc.} to users based on their personalized interests. Traditional RSs can be mainly categorized into three classes \cite{ricci2011introduction}: Collaborative filtering-based methods \cite{koren2022advances}, content-based methods \cite{lops2011content}, and hybrid methods \cite{ccano2017hybrid}. Collaborative filtering-based RSs estimate user interests and predict their future behaviors by exploiting their past activities, such as browsing, clicking, purchases, etc. Content-based methods, on the other hand, predict new recommendations by matching user interests with item content. Hybrid methods combine the advantages of both worlds, where collaborative information and user/item feature information are comprehensively considered to generate more accurate recommendations. 

Although recent years have witnessed substantial achievements for all three classes of RSs introduced above, a great limitation of these methods is that they can only estimate user interests and predict future recommendations based on correlations in the observational user historical behaviors and user/item features, which
guarantee no causal implications \cite{yao2021survey,gao2022causal}. For example, a collaborative filtering-based RS may discover that several drama shows from a certain genre \textit{tend to} have high ratings from a group of users, and conclude that we should keep recommending drama shows from the same genre to these users. But there is an important question: Are the high ratings caused by the fact that the users indeed like drama shows from this genre, or they were limitedly exposed to drama shows from the same genre (i.e., exposure bias), and if given a chance, they would prefer something new to watch? In addition, a content-based RS may observe that micro-videos with certain features \textit{are associated with} more clicks and conclude that these features may reflect the current trend of user interests. But are the clicks because these micro-videos tend to have sensational titles as clickbait where users could be easily deceived? Moreover, if the titles of these micro-videos are changed to the ones that reflect their true content, would users still click them? The above questions are causal in nature because they either ask about the effects of an intervention (e.g., what the rating would be if a new drama show \textbf{is made exposed} to the user) or a counterfactual outcome (e.g., would the user still click a micro-video if its title \textbf{had been changed} to faithfully reflect the content), rather than mere associations in the observational data. According to Pearl \cite{pearl2018book}, these questions lie on Rungs 2 and 3 of the Ladder of Causality, i.e., interventional and counterfactual reasoning, and they cannot be answered by traditional RSs that reason only with associations, which lie on Rung 1 of the ladder.

Why are these causal questions important for RSs? The first reason is that failing to address them may easily incur bias in recommendations, which can get unnoticed for a long time. If the collaborative filtering-based RSs mentioned above mistake exposure bias for user interests, they would amplify the bias by continuously recommending users with similar items; eventually, recommendations will lose serendipity, and users' online experience can be severely degraded. Similarly, for the content-based micro-video RSs, if they cannot distinguish clicks due to user interests from the ones deceived by clickbait, they may over-recommend micro-videos with sensational titles, which is unfair to the uploaders of high-quality micro-videos who put much effort into designing the content. In addition, understanding the cause of user activities can help improve the explainability of recommendations. Consider the causal question of whether a user purchases an item due to its quality or its low price. Pursuing the causal explanations behind user behaviors can help service providers to enhance the RS algorithm based on users' personalized preferences. Finally, causal inference allows us to identify and base recommendations on causal relations that are stable and invariant, while discarding other correlations that are undesirable or susceptible to change. Take restaurant recommendations as an example. Users can choose a restaurant because of its convenience (e.g., going to a nearby fast food shop to quickly grab a bite, but they do not necessarily like it, a non-stable correlation) or due to their personal interests (e.g., traveling far away for a hot-pot restaurant, a stable causal relation). If an RS can properly disentangle users' intent that causally affects their previous restaurant visits, even if the convenience levels of different restaurants may change due to various internal or external reasons such as users' moving to a new place, the system can still adapt well to the new situation. From this aspect, the generalization ability of the causal RSs can be substantially improved.

\begin{figure*}[t]
\centering
\includegraphics[width=.95\textwidth,]{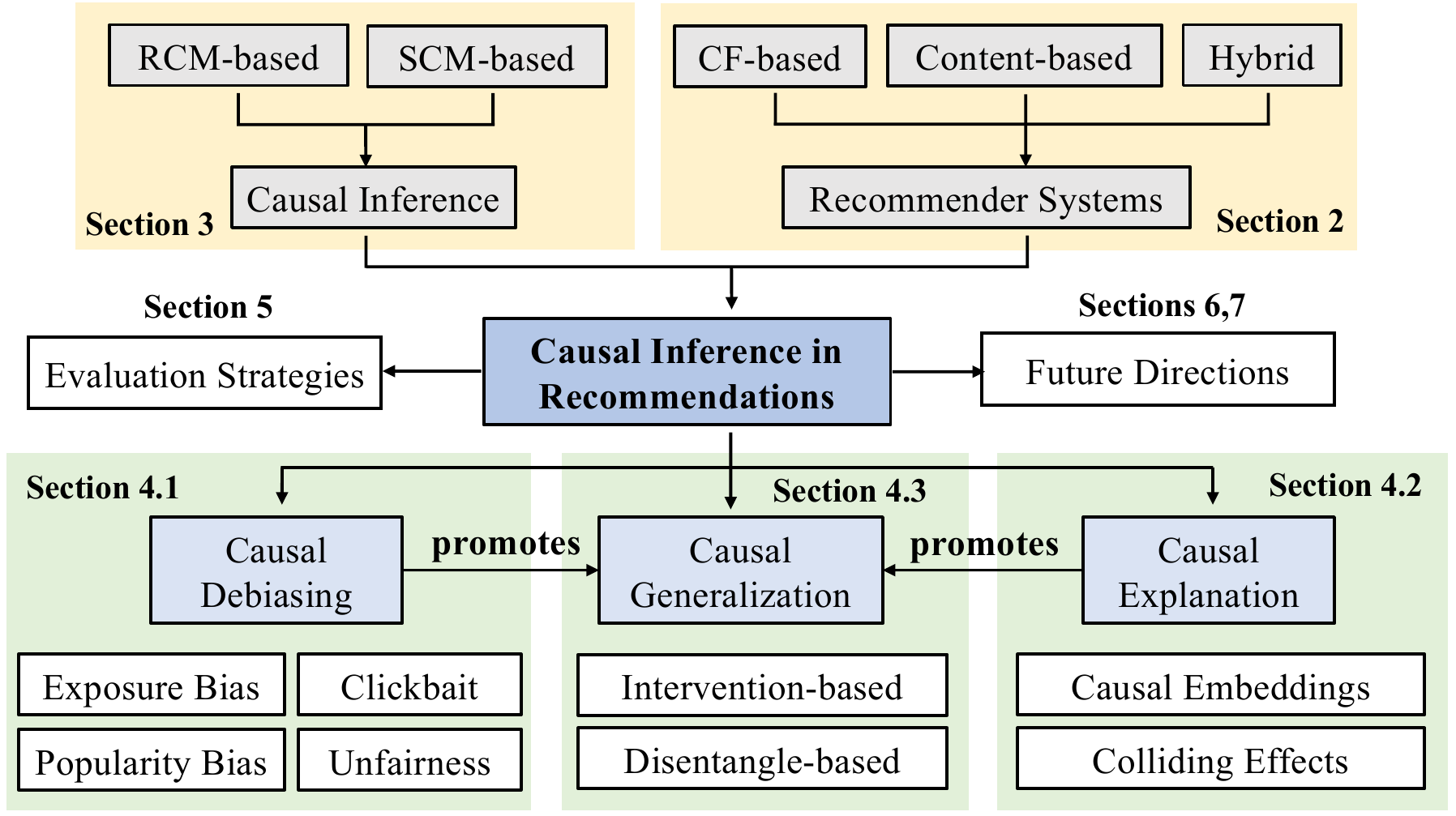}
      \caption{An overview of the structure of this survey and connections between different sections.}
       \label{fig:overview}
\end{figure*}

This survey provides a systematic overview of recent advances in causal RS research. The organization is illustrated in Fig. \ref{fig:overview}. We start with the fundamental concepts of traditional RSs and their limitation of correlational reasoning in Section \ref{sec:rs_basics}. Then Section \ref{sec:ci_recap} recaps two important causal inference paradigms in machine learning and statistics, and shows their connections with the recommendation task. Section \ref{sec:method} thoroughly discusses how different causal inference techniques can be introduced to address the limitations of traditional RSs, with an emphasis on debiasing, explainability promotion, and generalization improvement. Section \ref{sec:datasets} summarizes the general evaluation strategies for causal RSs. Finally, Sections \ref{sec:future} and \ref{sec:conclusions} discuss prospective open questions and future directions for causal RSs and conclude this survey.

\vspace{-2mm}

\section{Recommender System Basics}
\label{sec:rs_basics}

To keep this survey compact, we confine our discussions to simple RSs with $I$ users and $J$ items. The main data for the RSs, i.e., users' historical behaviors, are represented by a user-item rating matrix $\mathbf{R} \in \mathbb{R}^{I \times J}$, where a non-zero element $r_{ij}$ denotes user $i$'s rating to item $j$, and a zero element $r^{0}_{ik}$ indicates the rating is missing \footnote{We use rating to refer to any user-item interaction that can be represented by a numerical value. This includes both explicit feedback such as likes/dislikes, and implicit feedback such as views and clicks. When $r_{ij}$ represents implicit feedback, the missing elements $r^{0}_{ik}$ in $\mathbf{R}$ may be used as weak negative feedback in the training phase \cite{hu2008collaborative}. This may complicate the causal problems. Therefore, we assume RSs are trained on observed ratings to simplify the discussion unless specified otherwise.}. To make the discussions of RSs compatible with causal inference, we take a probabilistic view of $\mathbf{R}$ \cite{mnih2007probabilistic}, where $r_{ij}$ is assumed to be the realized value of the random variable $R$ dependent on user $i$ and item $j$\footnote{However, we do not distinguish random variables and their specific realizations if there is no risk of confusion. For simplicity, we assume $R$ to be Gaussian unless specified otherwise.}. In addition to $\mathbf{R}$, an RS usually has access to side information like user features $\mathbf{f}^{u}_{i} \in \mathbb{R}^{K^{u}_{F}}$, such as her age, gender, location, etc., or item features $\mathbf{f}^{v}_{j} \in \mathbb{R}^{K^{v}_{F}}$, such as its content and textual description. $K^{u}_{F}$ and $K^{v}_{F}$ are the dimensions of user and item features, respectively. \textbf{The main purpose of an RS} is to predict users' ratings for previously uninteracted items (i.e., the missing values $r^{0}_{ik}$ in $\mathbf{R}$) based on the observed ratings $r_{ij}$ in $\mathbf{R}$ and the available user and item side information such as $\mathbf{f}^{u}_{i}$ and $\mathbf{f}^{v}_{j}$, such that new relevant items can be properly recommended based on users' personalized interests. 


\subsection{Collaborative Filtering}
\label{sec:rscf}
Collaborative filtering-based RSs recommend new items by leveraging user ratings in the past. They generally consider the ratings $r_{ij}$ as being generated from a user latent variable $\mathbf{u}_{i} \in \mathbb{R}^{K}$ that represents user interests and an item latent variable $\mathbf{v}_{j} \in \mathbb{R}^{K}$ that encodes the item attributes (i.e., item latent semantic information), where $K$ is the dimension of the latent space. Here we list three widely-used collaborative filtering-based RSs, which will be frequently used as examples in this survey:
\begin{itemize}[parsep=6pt]
    \item \textbf{Matrix Factorization (MF)} \cite{koren2009matrix}. MF models $r_{ij}$ with the inner product between $\mathbf{u}_{i}$ and $\mathbf{v}_{j}$, where $r_{ij} \sim \mathcal{N}(\mathbf{u}_{i}^{T} \cdot \mathbf{v}_{j}, \sigma_{ij}^{2})$ and $\sigma_{ij}^{2}$ is the predetermined variance\footnote{For works that do not explicitly treat $r_{ij}$ as a random variable, we assume it follows a Gaussian distribution with zero variance. The generative process then becomes as $r_{ij} = \mathbf{u}_{i}^{T} \cdot \mathbf{v}_{j}$. }.

    \item \textbf{Deep Matrix Factorization (DMF) \cite{xue2017deep}}. DMF extends MF by applying deep neural networks (DNNs) \cite{zhang2019deep}, i.e., $f^{u}_{nn}, f^{v}_{nn}: \mathbb{R}^{K} \rightarrow \mathbb{R}^{K'}$, to $\mathbf{u}_{i}$ and $\mathbf{v}_{j}$, where the ratings are assumed to be generated as $r_{ij} \sim \mathcal{N}(f_{nn}^{u}(\mathbf{u}_{i})^{T} \cdot f^{v}_{nn}(\mathbf{v}_{j}), \sigma_{ij}^{2})$.
    
    \item  \textbf{Auto-encoder (AE)-based RSs \cite{wu2016collaborative,liang2018variational}} model user $i$'s ratings to all $J$ items as $\mathbf{r}_{i} \sim \mathcal{N}(f^{u}_{nn}(\mathbf{u}_{i}), \boldsymbol{\sigma}_{i}^{2} \cdot \mathbf{I}_{K})$, where $f^{u}_{nn}: \mathbb{R}^{K} \rightarrow \mathbb{R}^{J}$ is a DNN and item latent variables $\mathbf{v}_{j}$ for all items are implicit in last layer weights of the decoder \cite{zhu2022mutually}. 
\end{itemize}
In the training phase, the models learn the latent variables $\mathbf{u}_{i}$, $\mathbf{v}_{j}$ and the associated function $f_{nn}$ by fitting on 
 the \textbf{observed ratings} $r_{ij}$ (e.g., via maximum likelihood estimation, which essentially estimates the conditional distribution $p(r_{ij}|\mathbf{u}_{i}, \mathbf{v}_{j})$ from the observational data \cite{xu2021causal}). Afterward, we can use them to predict new ratings for previously uninteracted items $k$, e.g., $\hat{r}^{\textrm{MF}}_{ik} = \mathbf{u}_{i}^{T} \cdot \mathbf{v}_{k}$ for MF, $\hat{r}^{\textrm{DMF}}_{ik} = f_{nn}^{u}(\mathbf{u}_{i})^{T} \cdot f^{v}_{nn}(\mathbf{v}_{k})$ for DMF, and $\hat{r}^{\textrm{AE}}_{ik} = f^{u}_{nn}(\mathbf{u}_{i})_{k}$ for AE-based RSs, where the top ones that best match users' interests can be selected as recommendations. 
\begin{svgraybox}\textbf{Traditional collaborative filtering-based RSs reasons with correlations.} Ideally, we would expect $\mathbf{u}_{i}$, $\mathbf{v}_{j}$ and $f_{nn}$ to capture the causal influence of user interests and item attributes on ratings, i.e., what the rating would be if item $j$ is made exposed to user $i$ \cite{xu2021causal}. However, since the collected rating data are \textbf{observational} rather than experimental, what actually learned by $\mathbf{u}_{i}$, $\mathbf{v}_{j}$, and $f$ are the co-occurrence patterns in users' past behaviors, which guarantee no causal implications. Consequently, spurious correlations and biases can be captured by the model, which will be amplified in future recommendations \cite{wang2021deconfounded}. Furthermore, the learned user latent variable $\mathbf{u}_{i}$ generally entangles different factors that causally determine user interests. From this perspective, the explainability and generalization of these methods cannot be guaranteed.\end{svgraybox}

\subsection{Content-Based Recommender Systems}
\label{sec:cbrs}

Personalized content-based RSs (CBRSs) estimate user interests based on the features of the items they have interacted with. These models typically encode user interests into user latent variables $\mathbf{u}_{i} \in \mathbb{R}^{K}$ and assume that the ratings are generated by matching user interests with item content, i.e., $r_{ij} \sim \mathcal{N}(f(\mathbf{u}_{i}, \mathbf{f}^{v}_{j}), \sigma_{ij})$, where $f$ is a matching function. The training of personalized CBRSs follow similar steps as collaborative filtering, where $\mathbf{u}_{i}$ and $f$ are learned by fitting on the \textbf{observed ratings} (which essentially estimates the conditional distribution $p(r_{ij}|\mathbf{u}_{i}, \mathbf{f}^{v}_{j})$ from the observational data), and new ratings can be predicted by $\hat{r}_{ik} = f(\mathbf{u}_{i}, \mathbf{f}^{v}_{k})$. The key step of building a CBRS is to create item features $\mathbf{f}^{v}_{j}$ that can best reflect user interests, which crucially depends on the item being recommended. For example, for micro-videos, the visual, audio, and textual modalities are comprehensively considered such that users' interest in different aspects of a micro-video can be well captured \cite{wei2019mmgcn}. 

\begin{svgraybox}
\textbf{Traditional content-based RSs cannot model the causal influence of user interests $\mathbf{u}_{i}$ and item content $\mathbf{f}^{v}_{j}$ on user rating $r_{ij}$.} The reason is that, factors other than users' interests in the item content, such as users' being deceived by clickbaits (e.g., sensational titles of micro-videos) \cite{wang2021clicks}, etc., can create an undesirable association between item content $\mathbf{f}^{v}_{j}$ and user ratings $r_{ij}$ in the observed dataset, where the bias can be captured by the user latent variables $\mathbf{u}_{i}$ and the matching function $f$, and perpetuates into future recommendations.\end{svgraybox}

\vspace{-7mm}

\subsection{Hybrid Recommendation}

Hybrid RSs combine user/item side information with collaborative filtering to enhance the recommendations. A commonly-used hybrid strategy is to augment user and item latent variables $\mathbf{u}_{i}$ and $\mathbf{v}_{j}$ with user/item side information $\mathbf{f}^{v}_{i}$ and $\mathbf{f}^{v}_{j}$ in existing collaborative filtering methods by replacing $\mathbf{u}_{i}$ and $\mathbf{v}_{j}$ with $\mathbf{u}^{+}_{i} = [\mathbf{u}_{i} || \mathbf{f}^{u}_{i}]$ and $\mathbf{v}^{+}_{j} = [\mathbf{v}_{j} || \mathbf{f}^{v}_{j}]$ in MF, DMF, and AE-based RSs, where $[\cdot || \cdot]$ represents vector concatenation \cite{koren2008factorization,zhu2022variational}. The dimensions of $\mathbf{u}_{i}$ and $\mathbf{v}_{j}$ that encode the collaborative information are adjusted accordingly to make $\mathbf{u}^{+}_{i}$ and $\mathbf{v}^{+}_{j}$ compatible in the model. Another important class of hybrid RS is the factorization machine (FM) \cite{rendle2010factorization} and its extensions like \cite{he2017neuralf,li2022causalfm}, which can be viewed as learning a bi-linear function $f_{fm}$ where the ratings are generated by  $r_{ij} \sim \mathcal{N}(f_{fm}(\mathbf{u}_{i},\mathbf{v}_{j},\mathbf{f}^{u}_{i}, \mathbf{f}^{v}_{j}), \sigma_{ij}^{2})$. 
\begin{svgraybox}
\textbf{Simple hybrid strategies cannot break the correlational reasoning limitation of collaborative filtering and content-based RSs}, because the objective of the hybridization is still to improve the models' fitting on the observational user historical behaviors (i.e., estimating conditional distribution $p(r_{ij}|\mathbf{u}_{i},\mathbf{v}_{j},\mathbf{f}^{u}_{i}, \mathbf{f}^{v}_{j})$ from the data), where the causal reasons that lead to the observed user behaviors
are not considered. However, the idea of introducing extra user/item side information is important for building causal RSs. The reason is that, combined with the domain knowledge of human experts, the side information can help form more comprehensive causal relations among the variables of interests, such as user interests, item attributes, historical ratings, and other important covariates that may lead to spurious correlations and biases, which is usually a crucial step for causal reasoning in recommendations.
\end{svgraybox}

\section{Causal Recommender Systems: Preliminaries}
\label{sec:ci_recap}

In the previous section, we discussed the recommendation strategies of the traditional RSs and their limitations due to correlational reasoning on observational user behaviors. In this section, we introduce two causal inference frameworks, i.e., Rubin's potential outcome framework (also known as the Rubin causal model, RCM) \cite{imbens2015causal} and Pearl's structural causal model (SCM) \cite{pearl2009causality}, in the context of RSs, aiming to provide a theoretically rigorous basis for reasoning with correlation and causation in recommendations. We show that both RCM and SCM are powerful frameworks to build RSs with causal reasoning ability (i.e., causal RSs), but they are best suited for different tasks and questions. The discussions in this section serve as the foundation for more in-depth discussions of the state-of-the-art causal RS models in Section \ref{sec:method}.

\vspace{-3.5mm}

\subsection{Rubin's Potential Outcome Framework}
\label{sec:RCM}

\subsubsection{Motivation of Applications in RSs}
\label{sec:motiRCM}
To understand the correlational reasoning nature of traditional RSs, we note that naively fitting models on the observed ratings can only answer the question “what the rating would be \textbf{if we observe an item was exposed to the user}". Since item exposure is not randomized in the collected dataset \footnote{which can be attributed to multiple reasons such as users' self-search \cite{wang2021combating}, the recommendations of previous models \cite{liu2020general}, the position where the items are displayed \cite{wang2018position}, item popularity \cite{abdollahpouri2017controlling}, etc. Generally, RCM-based causal RSs are agnostic to the specific reason that causes the exposure bias.}, the predicate “the item was exposed to the user" \textit{per se} contains extra information about the user-item pair (e.g., the item could be more popular than other non-exposed items), which cannot be generalized to the rating predictions of \textbf{arbitrary} user-item pairs. Therefore, what RS asks is essentially an interventional question (and therefore a causal inference question), i.e., what the rating would be \textbf{if an item is made exposed to the user}. To address this question, RCM-based RSs draw inspiration from clinical trials, where exposing a user to an item is compared to subjecting a patient to a treatment, and the user ratings are analogous to the outcomes of the patients after the treatment \cite{schnabel2016recommendations,wu2022opportunity}. Accordingly, RCM-based RSs aim to estimate the causal effects of the treatments (exposing a user to an item) on the outcomes (user ratings), despite the possible correlations between the treatment assignment and the outcome observations \cite{schnabel2016recommendations}.

\subsubsection{Definitions and Objectives}

We first introduce necessary symbols and definitions to connect RCM with RSs. We consider the unit as the user-item pair $(i,j)$ that can receive the treatment “exposing user $i$ to item $j$”, and the population as all user-item pairs $\mathcal{PO} = \{(i,j), 1 \leq i, j \leq I, J\}$ \cite{bonner2018causal}. We start by using a binary scalar $a_{ij}$ to denote the exposure status of item $j$ for user $i$, i.e., the assigned treatment. We further define the \textbf{rating potential outcome} $r_{ij}(a_{ij}=1)$ as user $i$'s rating to item $j$ if the item is made exposed to the user and $r_{ij}(a_{ij}=0)$  as the rating if the item is not exposed \cite{wang2020causal}. For user $i$, if she rated item $j$, we observe $r_{ij}(a_{ij}=1) = r_{ij}$. Otherwise, we observe the baseline potential outcome $r_{ij}(a_{ij}=0)=0$, which is usually ignored in debias-oriented causal RS research \cite{schnabel2016recommendations,steck2010training}\footnote{In the uplift evaluation of RSs that aims to estimate how recommendations change user behaviors \cite{sharma2015estimating}, $r_{ij}(a_{ij}=0)$ may be used to represent user $i$'s rating to item $j$ through self-searching \cite{sato2019uplift}.}. Similar to clinical trials, we can define the treatment group $\mathcal{T} = \{(i,j):a_{ij}=1\}$ as the set of user-item pairs where user $i$ is exposed to item $j$, and define the non-treatment group $\mathcal{NT} = \{(i,k) :a_{ik}=0\}$ accordingly. \textbf{The purpose of RSs, under the RCM framework}, can be framed as utilizing the observed ratings from units in the treatment group $\mathcal{T}$ to unbiasedly estimate the rating potential outcomes for units from the population $\mathcal{PO}$, despite the possible correlations between item exposures $a_{ij}$ and user ratings $r_{ij}$ in the collected data.

\subsubsection{Causal Analysis of Traditional RSs}
\label{sec:randomized_rcm}
\begin{figure*}[t]
\centering
\includegraphics[width=.78\textwidth,]{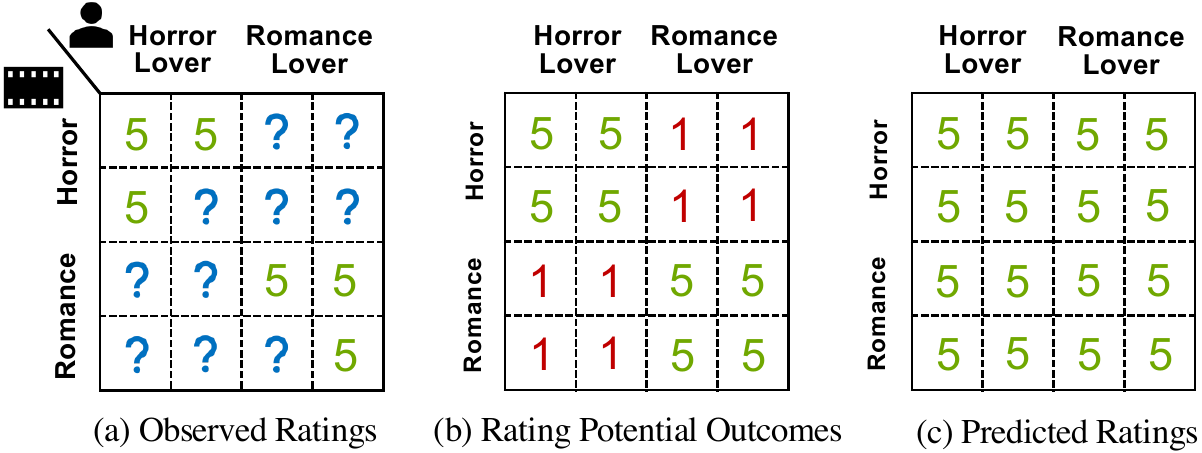}
      \caption{A classical example of exposure bias in RSs \cite{steck2010training}. The example is composed of two horror lovers who always rate horror movies with five while hating romance movies, and two romance lovers would who do exactly the opposite. (a) shows the observed ratings $r_{ij}$. (b) shows the rating potential outcomes $r_{ij}(a_{ij}=1)$. (c) shows the rating predictions of an RS that maximizes the likelihood of the observed ratings in (a), but the RS is bad because it predicts all ratings to five.}
       \label{fig:RCM}
\end{figure*}

Traditional RSs naively train a rating prediction model that best fits the ratings in the treatment group $\mathcal{T}$ (e.g., via maximum likelihood introduced in Section \ref{sec:rs_basics}) to estimate the unobserved rating potential outcomes  $r_{ij}(a_{ij}=1)$ for user-item pairs in $\mathcal{NT}$ \cite{chen2020bias}, which neglect the fact that exposure bias can lead to a systematic difference in the distribution of $r_{ij}(a_{ij}=1)$ between $\mathcal{T}$ and $\mathcal{NT}$. For example, users tend to rate items they like in reality, which could lead to the following spurious correlation between item exposure $a_{ij}$ and rating potential outcome $r_{ij}(a_{ij}=1)$:
\begin{equation}
\label{eq:bias_rubin}
p(r_{ij}(a_{ij}=1) \mathrm{\ is\ high} | a_{ij}=1) > p(r_{ij}(a_{ij}=1) \mathrm{\ is\ high} | a_{ij}=0),   
\end{equation}
i.e., users who have rated an item $j$ may have systematically higher ratings than users who haven't rated it yet. In this case, traditional RSs may have a tendency to overestimate the ratings for units in $\mathcal{NT}$ (see Fig. \ref{fig:RCM} for an intuitive example). Theoretically, RCM attributes the exposure bias in the collected dataset to the violation of the \textbf{unconfoundedness assumption} \cite{imbens2015causal} defined as follows:
\begin{equation}
\label{eq:si_assu}
    r_{ij}(a_{ij}=1) \perp a_{ij}.
\end{equation}
The rationale is that, if Eq. (\ref{eq:si_assu}) holds, the exposure of user $i$ to item $j$ (i.e., $a_{ij}$) is independent of the rating potential outcome $r_{ij}(a_{ij}=1)$, which implies that $r_{ij}(a_{ij}=1)$ in $\mathcal{T}$ and $\mathcal{NT}$ follows the same distribution. Therefore, the exposure of the items is randomized, and exposure bias such as Eq. (\ref{eq:bias_rubin}) will not exist \cite{wang2020causal}.  

\subsubsection{Potential Outcome Estimation with the RCM Framework}
One classic solution from the RCM-based framework to address the exposure bias is that we find user and item covariates $C$, such that in each data stratum specified by $C=\mathbf{c}$, users' exposure to items are randomized \cite{imbens2015causal}. The property of the covariates $C$ can be formulated as the conditional unconfoundedness assumption as follows:
\begin{equation}
\label{eq:csi_assu}
    r_{ij}(a_{ij}=1) \perp a_{ij} \mid  \mathbf{c}.
\end{equation}
$C$ is sometimes non-rigorously referred to as \textbf{confounder} in the literature, but we will see its formal definition in the next subsection. If Eq. (\ref{eq:csi_assu}) holds, the item exposures are independent of the rating potential outcomes in each data stratum specified by $C=\mathbf{c}$, and the exposure bias can be attributed solely to the discrepancy in the distribution of the covariates $C=\mathbf{c}$ between the treatment group $\mathcal{T}$ and the population $\mathcal{PO}$, i.e.,  $p(\mathbf{c}|a_{ij}=1)$ and $p(\mathbf{c})$\footnote{{\color{red}!} We can gain an intuition of this claim from Fig. \ref{fig:RCM}. Suppose covariates $C$ represent the two-dimensional features (user type, movie type). Given $C=\mathbf{c}$, $ r_{ij}(a_{ij}=1) \perp a_{ij} \mid  \mathbf{c}$ described in Eq. (\ref{eq:csi_assu}) is satisfied because in each data stratum specified by $C=\mathbf{c}$ (i.e., the four $2 \times 2$ blocks in Fig. \ref{fig:RCM}-(b)), $r_{ij}(a_{ij}=1)$ is constant. Fig. \ref{fig:RCM}-(a) shows that for the treatment group $\mathcal{T}$, $p(\mathbf{c}|a_{ij}=1)=1/2$ for $\mathbf{c} \in \mathcal{C}_{1} = \{(\text{horror fan, horror movie}), (\text{romance fan, romance movie})\}$ and $p(\mathbf{c}|a_{ij}=1)=0$ for $\mathbf{c} \in \mathcal{C}_{2} = \{(\text{horror fan, romance movie}), (\text{romance fan, horror movie})\}$. In contrast, for the population $\mathcal{PO}$, $p(\mathbf{c})=1/4$ for $\mathbf{c} \in \mathcal{C}_{1} \cup \mathcal{C}_{2}$. Therefore, in the treatment group $\mathcal{T}$, user-item pairs with covariates in $\mathcal{C}_{1}$ are over-represented, while those with covariates in $\mathcal{C}_{2}$ are under-represented. However, we also note that this case is too extreme to be addressed by RCM, as $p(\mathbf{c}|a_{ij}=1)=0$ for $C \in \mathcal{C}_{2}$ \textbf{violates} the positivity assumption mentioned in the above attention box.} Therefore, we can reweight the observed ratings in $\mathcal{T}$ based on the covariates $C$ to address the bias, such that they can be viewed as pseudo randomized samples. This leads to inverse propensity weighting (IPW), which eliminates the exposure bias from the data's perspective \cite{schnabel2016recommendations}. In addition, we can also adjust the influence of $C$ in the RS model, where the exposure bias is addressed from the model side \cite{wang2020causal}. Both methods will be discussed in Section \ref{sec:exp_bias}.

\begin{warning}{Attention: Extra Assumptions Required by Most RCM-based RSs}In addition to unconfoundedness, most RCM-based RS need two extra assumptions to identify the causal effects of item exposures on ratings: (1) The \textbf{stable unit treatment assumption (SUTVA)}, which states that items exposed to one user cannot affect ratings of another user. (2) The \textbf{positivity assumption}, which states that every user has a positive chance of being exposed to every item \cite{imbens2015causal}. For RCM-based causal RSs introduced in this survey, these two assumptions are tacitly accepted.
\end{warning}

\subsection{Pearl's Structural Causal Model}
\subsubsection{Motivation of Applications in RS}
Different from RCM that uses rating \textit{potential outcomes} to reason with causality and attributes the biases in observed user behaviors to non-randomized item exposures, Pearl's structural causal model (SCM) delves deep into the causal mechanism that generates the \textit{observed outcomes} (and  biases) and represents it with a causal graph $G = (\mathcal{N}, \mathcal{E})$. The nodes $\mathcal{N}$ specify the variables of interests, which in the context of RS could be user interests $U$, item attributes $V$, observed ratings $R$, and other important covariates $C$, such as item popularity, user features, etc\footnote{In causal graphs, the subscripts $i$, $j$ for each node are omitted for simplicity.}. The directed edges $\mathcal{E}$ between nodes represent their causal relations determined by researchers' domain knowledge. Each node $X \in \mathcal{N}$ is associated with a structural equation $p_{G}(X|Pa(X))$\footnote{We also omit the mutually independent exogenous variables for each node and summarize their randomness into the structural equations with probability distributions \cite{gao2022causal}. Subscript $G$ is used to distinguish structural equations from other conditional relationships that can be inferred from $G$.}, which describes how the parent nodes $Pa(X)$ causally influence $X$ (i.e., the response of $X$ when setting nodes in $Pa(X)$ to specific values)

Although RCM and SCM are generally believed to be fundamentally equivalent \cite{pearl2009causality}, both have their unique advantages. Compared to RCM, the key advantage of SCM is that causal graph offers an intuitive and straightforward way to encode and communicate domain knowledge and substantive assumptions of researchers, which is beneficial even for the RCM-based RSs \cite{wang2020causal}. Furthermore, SCM is more flexible as it can represent and reason with the causal effects between any subset of nodes in the causal graph (e.g., between two causes $U, V$ and one outcome $R$), as well as the causal effects along specific paths (e.g., $U \rightarrow R$ and $U_{c} \rightarrow R$). Therefore, SCMs are broadly applicable to multiple problems in RSs (not limited to exposure bias), such as clickbait bias, unfairness, entanglement, domain adaptation, etc \cite{gao2022causal}.

\begin{warning}{Attention: Two Caveats of SCM-based Causal RSs.}There are two caveats of SCM-based causal RSs. (1) Causal graphs for RSs often involve user, item \textbf{latent variables} $U$, $V$ that encode user interests and item attributes. Most works infer them alongside the estimation of structural equations and treat them as if they were observed when analyzing the causal relations. Alternatively, this can be viewed as representing users and items with their IDs (i.e., $i$ and $j$) in the causal graph and subsuming the embedding process into the structural equations \cite{xu2022dynamic}. (2) Generally, the causal graph should describe the causal mechanism that generates the \textbf{observed data}, because it allows us to distinguish invariant, causal relations from undesirable correlations. For example, we may argue that item popularity $C$ should be determined by item attributes $V$, i.e., $V \rightarrow C$. But to describe the generation of the observed ratings, causal relation $C \rightarrow V$ is usually assumed instead as item popularity causally influences the exposure probability of each item \cite{zhang2021causal}.
\end{warning}

\subsubsection{Atomic Structures of Causal Graphs}
\label{sec:atom}
\begin{figure*}[t]
\centering
\includegraphics[width=.86\textwidth,]{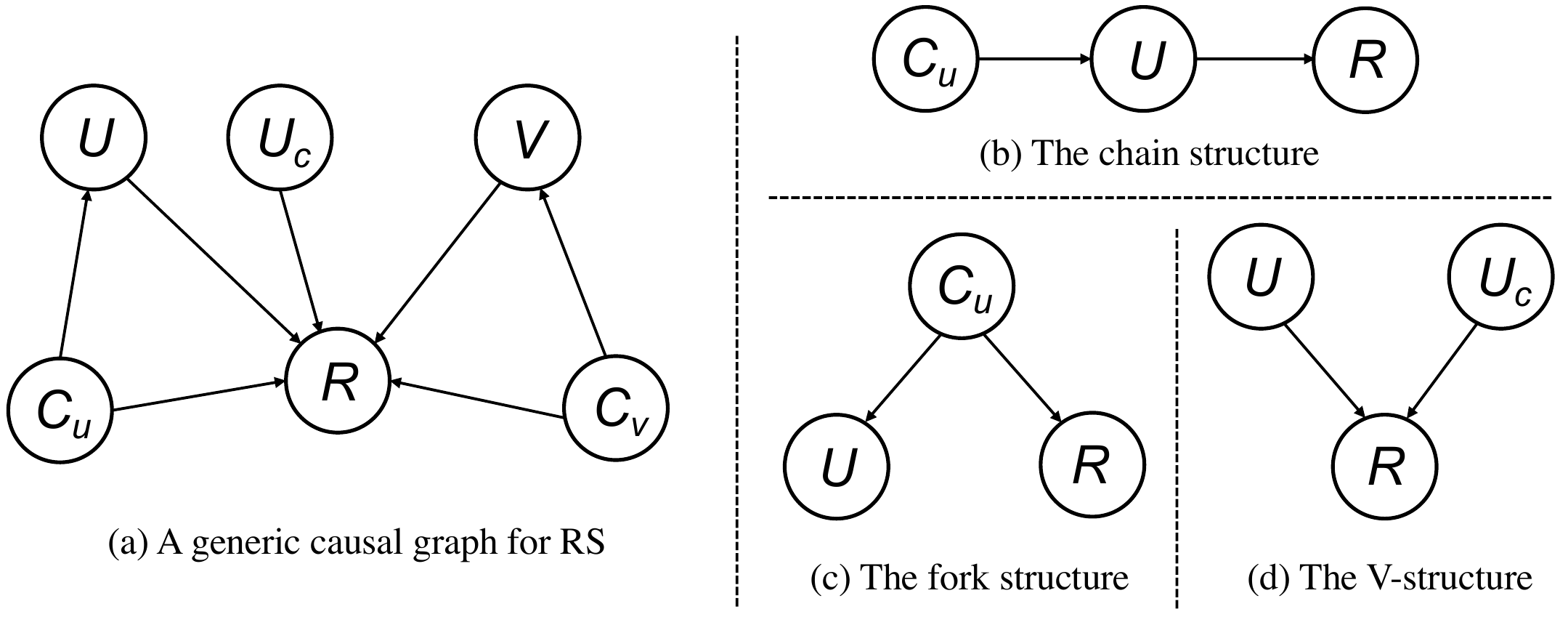}
      \caption{(a): A generic causal graph for RS that depicts the causal influence of user interests $U$, user conformity to the popularity trend $U_{c}$, and item attributes $V$ on the observed ratings $R$. Specifically, the causal paths $U \rightarrow R$ and $V \rightarrow R$ are confounded by $C_{u}$ and $C_{v}$, which represent user features and item popularity, respectively. (b)(c)(d): Three atomic structures identified from (a).}
       \label{fig:scm-overview}
\end{figure*}

The structure of causal graphs represents researchers' domain knowledge regarding the causal generation process of the observational data, which is the key to distinguishing stable, causal relations from other undesirable correlations between variables of interest. Here, we use a generic causal graph applicable to RSs in Fig. \ref{fig:scm-overview}-(a) as a running example to illustrate three atomic graph structures:

\begin{itemize}[parsep=6pt]
    \item \textbf{Chain,} e.g., $C_{u} \rightarrow U \rightarrow R$. In a chain, the successor node is assumed to be causally influenced by the ancestor nodes. In the example, $U$ is a direct cause of $R$, whereas $C_{u}$ indirectly influences $R$ via $U$ as a mediator.
    \item \textbf{Fork,} e.g., $U \leftarrow C_{u} \rightarrow R$. In the fork, $C_{u}$ is called a \textbf{confounder} as it causally influences two children $U$ and $R$. From a probabilistic perspective, $U$ and $R$ are \textbf{not} independent unless conditioned on the confounder $C_{u}$ \cite{koller2009probabilistic}. This leads to the tricky part of a fork structure, i.e., \textbf{confounding effect} \cite{pearl2009causality}, where an unobserved $C_{u}$ can lead to spurious correlations between $U$ and $R$.
    \item \textbf{V-structure}, e.g., $U \rightarrow R \leftarrow U_{c}$. In the V-structure, $R$ is called a \textbf{collider} because it is under the causal influence of two parents, i.e., $U$ and $U_{c}$. An interesting property of the V-structure is the colliding effects \cite{pearl2009causality}, where observing $R$ creates a dependence on $U$ and $U_{c}$, even if they are marginally independent.
\end{itemize}

Confounders can lead to non-causal dependencies among variables in the observational dataset. This could introduce bias in traditional RSs, where the confounding effects are mistaken as causal relations. Confounding bias is a generic problem in RSs \cite{xu2021causal}, which will be further analyzed in the following subsections. In addition, abstracted V-structure usually leads to the entanglement of causes, which could jeopardize the explainability of RSs. For example, a user's purchase of an item may be due to her interest, i.e., $U$, or her conformity to the popularity trend, i.e., $U_{c}$. Since most RSs summarize both into a user latent variable $U$, the V-structure $U \rightarrow R \leftarrow U_{c}$ is abstracted away, where the two causes of the purchase cannot be distinguished. 

\subsubsection{Causal Analysis of Traditional RSs}

\begin{figure*}[t]
\centering
\includegraphics[width=.94\textwidth,]{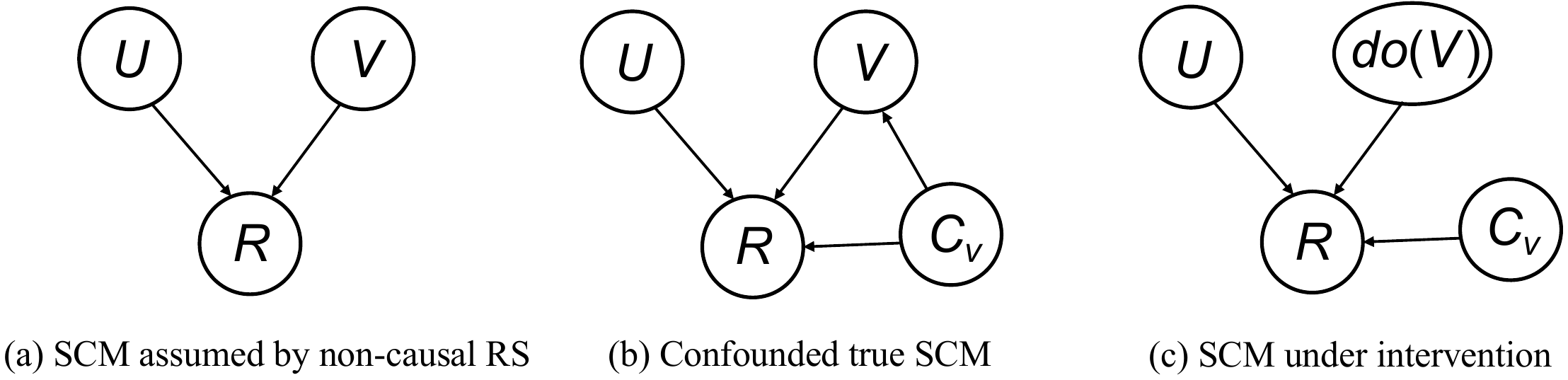}
      \caption{(a): SCM assumed by non-causal collaborative filtering-based RS. (b): The confounded SCM that depicts the true data generation process. (c): SCM under intervention $do(V)$.}
       \label{fig:scm-biasanalysis}
\end{figure*}

\vspace{-2mm}

In this section, we investigate the susceptibility of traditional collaborative filtering-based RSs to the confounding bias. As discussed in Section \ref{sec:rscf}, a commonality of these models is that they estimate conditional distribution $p(r_{ij}|\mathbf{u}_{i}, \mathbf{v}_{j})$ from observed ratings and use it to predict new ratings. For $p(r_{ij}|\mathbf{u}_{i}, \mathbf{v}_{j})$ to represent the causal influence of user interests $\mathbf{u}_{i}$ and item attributes $\mathbf{v}_{j}$ on ratings $r_{ij}$ (which, in the context of collaborative filtering, means the rating of any arbitrary item $j$ that is made exposed to user $i$ \cite{xu2021causal}), the causal graph $G_{1}$ of Fig. \ref{fig:scm-biasanalysis}-(a) is tacitly assumed, i.e., no unobserved confounders for causal paths $U \rightarrow R$ and $V \rightarrow R$\footnote{This corresponds to the case where item exposures are randomized (see the discussions in Section \ref{sec:randomized_rcm}), as the user-item pair $(U, V)$ is not determined by other factors associated with $R$ \cite{richardson2013single}.}. 


However, in reality, both $U \rightarrow R$ \cite{wang2021deconfounded,wei2021model} and $V \rightarrow R$  \cite{agarwal2019general,chen2021autodebias} can be confounded, where the confounding effects can be implicitly captured by $p(r_{ij}|\mathbf{u}_{i}, \mathbf{v}_{j})$ that bias future recommendations. To reveal the bias, we consider the scenario where the causal path $V \rightarrow R$ is confounded by $C_{v}$ (e.g., item popularity). We assume the causal path $C_{v} \rightarrow V$ denotes the causal influence of $C_{v}$ on the exposure probability of item $V$ \cite{zhang2021causal}. In this case, the observed ratings are generated according to the causal graph $G_{2}$ in Fig. \ref{fig:scm-biasanalysis}-(b). Utilizing the law of total probability, the conditional distribution $p(r_{ij}|\mathbf{u}_{i}, \mathbf{v}_{j})$ estimated from the confounded data can be calculated as:
\begin{equation}
\label{eq:bias_pearl}
    p(r_{ij}|\mathbf{u}_{i}, \mathbf{v}_{j}) = \sum_{\mathbf{c}} p(\mathbf{c}|\mathbf{v}_{j}) \cdot p_{G_{2}}(r_{ij}|\mathbf{u}_{i}, \mathbf{v}_{j}, \mathbf{c})= \mathbb{E}_{p(C_{v}|\mathbf{v}_{j})}[p_{G_{2}}(r_{ij}|\mathbf{u}_{i}, \mathbf{v}_{j}, C_{v})].
\end{equation}
The issue of Eq. (\ref{eq:bias_pearl}) is that, the $p(\mathbf{c}|\mathbf{v}_{j})$ term is not causal (as we only have an edge $C_{v} \rightarrow V$ in the causal graph but not $V \rightarrow C_{v}$). In fact, $p(\mathbf{c}|\mathbf{v}_{j})$ represents abductive reasoning because it infers the cause $\mathbf{c}$ (e.g., item popularity) from the effect $\mathbf{v}_{j}$ (i.e., item $j$ is exposed to user $i$) and uses the inferred $\mathbf{c}$ to support the prediction of the rating $r_{ij}$. However, such reasoning cannot be generalized to the rating prediction of an arbitrary item $\mathbf{v}_{j}$, i.e., an item that \textbf{is made exposed} to the user. In other words, uncontrolled confounder $C_{v}$ leaves open a \textbf{backdoor path} (i.e., non-causal path) between $V$ and $R$, such that non-causal dependence of $R$ on $V$ exists in the data, which can be captured by traditional RSs and bias future recommendations. \footnote{The similarity between this section and Section \ref{sec:motiRCM} shows us the connection between RCM-based and SCM-based causal RSs, where the claim that when item exposure is not randomized, “observing that an item was
exposed to the user \textit{per se} contains extra information about the user-item pair" is mathematically transformed into the abductive inference of $\mathbf{c}$ from $\mathbf{v}_{j}$ by $p(\mathbf{c}|\mathbf{v}_{j})$.} 

\subsubsection{Causal Reasoning with SCM}

To calculate the causal effect of $\mathbf{u}_{i}$ and $\mathbf{v}_{j}$ on $r_{ij}$, we should conduct \textbf{intervention} on $U$ and $V$. This means that we set $U$, $V$ to $\mathbf{u}_{i}$, $\mathbf{v}_{j}$ regardless of the values of their parent nodes in the causal graph, including the confounder $C_{v}$ (because these nodes determine the exposure of item $j$ to user $i$ in the observed data). SCM denotes the intervention with \textbf{do-operator} as $p(r_{ij}|do(\mathbf{u}_{i}, \mathbf{v}_{j}))$ to distinguish it from the conditional distribution $p(r_{ij}|\mathbf{u}_{i}, \mathbf{v}_{j})$ that reasons with correlations in the observational data. Consider again the causal graph $G_2$ illustrated in Fig. \ref{fig:scm-biasanalysis}-(b). The intervention on node $V$ can be realized by removing all the incoming edges for node $V$ and setting the structural equation $p_{G_{2}}(V|C_{v})$ deterministically as $V=\mathbf{v}_{j}$, while other structural equations remain intact (Fig. \ref{fig:scm-biasanalysis}-(c)). If the confounder $C_{v}$ can be determined and measured for each item, the interventional distribution $p(r_{ij}|do(\mathbf{u}_{i}, \mathbf{v}_{j}))$ can be directly calculated from the confounded data via \textbf{backdoor adjustment} \cite{pearl2009causality} as:  
\begin{equation}
\label{eq:bkd_adj}
    p(r_{ij}|do(\mathbf{u}_{i}, \mathbf{v}_{j})) = \sum_{\mathbf{c}} p_{G_{2}}(\mathbf{c}) \cdot p_{G_{2}}(r_{ij}|\mathbf{u}_{i}, \mathbf{v}_{j}, \mathbf{c})= \mathbb{E}_{p_{G_{2}}(C_{v})}[p_{G_{2}}(r_{ij}|\mathbf{u}_{i}, \mathbf{v}_{j}, C_{v})],
\end{equation}
which, compared with Eq. (\ref{eq:bias_pearl}), blocks the abductive inference of $\mathbf{c}$ from $\mathbf{v}_{j}$, such that the causal influence of $\mathbf{u}_{i}$, $\mathbf{v}_{j}$ on $r_{ij}$ can be properly identified. 

Backdoor adjustment requires all confounders to be determined and measured in advance, but there are other SCM-based causal inference methods that can estimate causal effects with unknown confounders, and we refer readers to \cite{xu2021deconfounded,zhu2022mitigating} for details. Moreover, causal graphs allow us to conduct other types of causal reasoning based on the encoded causal knowledge, such as debiasing for non-confounder-induced biases (e.g., clickbait bias and unfairness), causal disentanglement, and causal generalization \cite{zheng2021disentangling}. These will be thoroughly discussed in the next section.

\vspace{-3mm}

\section{Causal Recommender Systems: The State-of-the-Art}
\label{sec:method}

Based on the preliminary knowledge of RSs and causal inference discussed in previous sections, we are ready to introduce the state-of-the-art causal RSs. Specifically, we focus on three important topics, i.e., bias mitigation, explainability promotion, and generalization improvement, as well as their inter-connections, where various limitations of traditional RSs due to correlational reasoning can be well addressed.

\subsection{Causal Debiasing for Recommendations}

The correlational reasoning of traditional RSs can inherit multiple types of biases in the observational user behaviors and amplify them in future recommendations \cite{chen2020bias}. The biases may result in various consequences, such as the discrepancy between offline evaluation and online metrics, loss of diversity, reduced recommendation quality, offensive recommendations, etc. Causal inference can distinguish stable causal relations from spurious correlations and biases that could negatively influence the recommendations, such that the robustness of recommendations can be improved.

\subsubsection{Exposure Bias}
\label{sec:exp_bias}

Exposure bias in RSs broadly refers to the bias in observed ratings due to non-randomized item exposures. From the RCM's perspective, exposure bias can be defined as the bias \textit{where users are favorably exposed to items depending on their expected ratings for them (i.e., rating potential outcomes)} \cite{steck2010training}. Exposure bias occurs due to various reasons, such as users' self-search or the recommendation of the previous RSs \cite{liu2020general}, which leads to the down-weighting of items less likely to be exposed to users. Since item exposures can be naturally compared with treatments in clinical trials, we discuss the debiasing strategies with the RCM framework.
\vspace{0.2cm}

\noindent \textbf{Inverse Propensity Weighting (IPW).}  IPW-based causal RSs aim to reweight the biased observed ratings $r_{ij}$ for user-item pairs in the treatment group, i.e., $\mathcal{T} = \{(i, j): a_{ij}=1\}$, to create pseudo randomized samples \cite{sato2020unbiased} for unbiased training of RS models that aim to predict the rating potential outcomes $r_{ij}(a_{ij}=1)$ for the population  $\mathcal{PO} = \{(i,j), 1 \leq i, j \leq I, J\}$. Intuitively, we can set the weight of $r_{ij}$ for units in $\mathcal{T}$ to be the inverse of item $j$'s exposure probability to user $i$, such that under-exposed items can be up-weighted and vice versa. If for each user-item pair, the covariates $\mathbf{c}$ that satisfy the conditional unconfoundedness assumption in Eq. (\ref{eq:csi_assu}) are available, the exposure probability $e_{ij}$ can be unbiasedly estimated from $\mathbf{c}$ via 
\begin{equation}
\label{eq:propensity}
e_{ij} = p(a_{ij}=1 | \mathbf{c}) = \mathbb{E}[a_{ij} | \mathbf{c}],  
\end{equation}
which is formally known as \textbf{propensity score} in causal inference literature \cite{rosenbaum1983central}.

\begin{figure*}[t]
\centering
\includegraphics[width=.95\textwidth,]{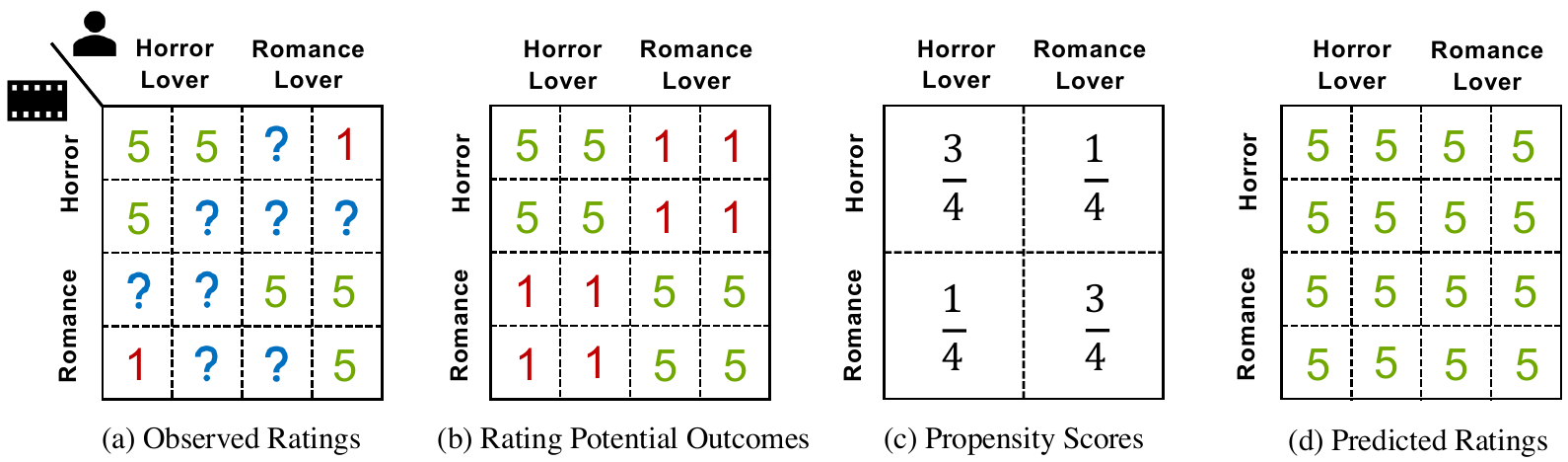}
      \caption{An example adapted from Fig. (\ref{fig:RCM}) where the positivity assumption holds. Suppose again covariates $C$ represent the two-dimensional features (user type, movie type). (a) shows the observed ratings; (b) shows rating potential outcomes; (d) shows the predicted rating potential outcome of an RS model. The propensity scores $e_{ij} = p(a_{ij}|\mathbf{c}) = \mathbb{E}[a_{ij} | \mathbf{c}]$ are shown in (c). Based on (a)(d) and Eq. (\ref{eq:before_ipw}), $\mathcal{L}^{\text{Obs}} = (5-1)^{2}\times 2 / 8 = 4$. Based on (b)(d) and Eq. (\ref{eq:True}), $L^{\text{True}} = (5-1)^{2}\times8/16 = 8$. Based on (a)(c)(d) and Eq. (\ref{eq:IPW}), $\mathcal{L}^{\text{IPW}} = \frac{1}{1/4}(5-1)^{2}\times2/16 = 8$, which is unbiased for $L^{\text{True}}$.
      }
       \label{fig:RCM_IPW}
\end{figure*}

\begin{important}{Background: The Balancing Property of Propensity Scores.}Propensity scores have the following property called balancing \cite{imbens2015causal, WinNT}, which is the key to proving the unbiasedness of IPW-based RSs:
\begin{equation}
\label{eq:blc_prop}
\begin{aligned}
    &\mathbb{E}\Big[\frac{r_{ij}}{e_{ij}} \Big| a_{ij}=1\Big] = \mathbb{E}\Big[\frac{r_{ij} \cdot a_{ij}}{e_{ij}} \Big] = \mathbb{E}\Big[\mathbb{E}\Big[\frac{r_{ij}\cdot a_{ij}}{e_{ij}} \Big| \mathbf{c}\Big]\Big]\\ 
    =\,\,& \mathbb{E}\Big[\mathbb{E}\Big[\frac{r_{ij}(a_{ij}=1)\cdot a_{ij}}{e_{ij}} \Big| \mathbf{c} \Big]\Big] \overset{(a)}{=} \mathbb{E}\Big[\frac{\mathbb{E}[r_{ij}(a_{ij}=1) \mid \mathbf{c}]\cdot \mathbb{E}[a_{ij} \mid \mathbf{c}]}{e_{ij}}\Big] \\
    =\,\,& \mathbb{E}\Big[\frac{\mathbb{E}[r_{ij}(a_{ij}=1) \mid \mathbf{c}]\cdot e_{ij}}{e_{ij}}\Big] = \mathbb{E}[r_{ij}(a_{ij}=1)],
\end{aligned}
\end{equation}
where the step $(a)$ follows the conditional unconfoundedness assumption in Eq. (\ref{eq:csi_assu}). 
\end{important}
We first discuss the implementation of IPW-based RS and its unbiasedness if user and item covariates $\mathbf{c}$ that satisfy Eq. (\ref{eq:csi_assu}) are available and the propensity scores $e_{ij}$ can be calculated exactly as Eq. (\ref{eq:propensity}). We denote the rating predictor of an RS that aims to predict the rating potential outcome $r_{ij}(a_{ij}=1)$ as $\hat{r}_{ij}$ and assume $r_{ij}(a_{ij}=1)$ follows the unit-variance Gaussian distribution. Ideally, we would like $\hat{r}_{ij}$ to maximize the log-likelihood on the rating potential outcomes $r_{ij}(a_{ij}=1)$ for all user-item pairs in $\mathcal{PO}$, which is equivalent to the minimization of the mean squared error (MSE) loss between $\hat{r}_{ij}$ and $r_{ij}(a_{ij}=1)$ as follows: 
\begin{equation}
\label{eq:True}
\mathcal{L}^{\text{True}} = \frac{1}{I \times J}\sum_{i,j}(\hat{r}_{ij} - r_{ij}(a_{ij}=1))^{2}.
\end{equation} However, since $r_{ij}(a_{ij}=1)$ is unobservable for user-item pairs in the non-treatment group $\mathcal{NT}$, $\mathcal{L}^{\text{True}}$ is impossible to calculate. Therefore, traditional RSs only maximize the log-likelihood of the observed ratings for user-item pairs in the treatment group $\mathcal{T}$, which leads to the empirical MSE loss as follows:
\begin{equation}
\label{eq:before_ipw}
   \mathcal{L}^{\text{Obs}} = \frac{1}{|(i,j):a_{ij}=1|} \sum_{(i,j):a_{ij}=1} (\hat{r}_{ij}-r_{ij})^{2},
\end{equation}
where $|(i,j):a_{ij}=1|$ is the number of observed ratings. When exposure bias exists, item exposure $a_{ij}$ depends on the rating potential outcome $r_{ij}(a_{ij}=1)$. Therefore, $\mathcal{L}^{\text{Obs}}$ is a biased estimator for $\mathcal{L}^{\text{True}}$, because the observed ratings for user-item pairs in the treatment group $\mathcal{T}$ are biased samples from the rating potential outcomes of the population $\mathcal{PO}$ (see Fig. \ref{fig:RCM_IPW}-(a) and Fig. \ref{fig:RCM_IPW}-(b) for an example). To remedy the bias, IPW-based causal RSs reweight the observed ratings $r_{ij}$ in $\mathcal{T}$ by the inverse of the propensity scores, i.e., $\frac{1}{e_{ij}}$, which leads to the following new training objective:
\begin{equation}
\label{eq:IPW}
    \mathcal{L}^{\text{IPW}} = \frac{1}{I \times J}\sum_{(i,j):a_{ij}=1} \frac{1}{e_{ij}} \cdot (\hat{r}_{ij}-r_{ij})^{2}.
\end{equation}
The proof for the unbiasedness of $\mathcal{L}^{\text{IPW}}$ for $\mathcal{L}^{\text{True}}$ can be achieved by utilizing the balancing property of propensity scores in Eq. (\ref{eq:blc_prop}), where we substitute $(\hat{r}_{ij}-r_{ij})^{2}$ for $r_{ij}$ in the LHS of Eq. (\ref{eq:blc_prop}) and treat the rating predictor $\hat{r}_{ij}$ as constant \cite{schnabel2016recommendations}. We also provide a toy example in Fig. \ref{fig:RCM_IPW} to intuitively show the calculation of $e_{ij}$, the biasedness of $\mathcal{L}^{\text{Obs}}$ and the unbiasedness of $\mathcal{L}^{\text{IPW}}$. The objective for IPW-based RSs defined in Eq. (\ref{eq:IPW}) is model-agnostic. Therefore, it is applicable to all traditional RSs we introduced in Section \ref{sec:rs_basics}. For example, for MF-based RSs, we can plug in $\hat{r}^{\textrm{MF}}_{ij} = \mathbf{u}_{i}^{T} \cdot \mathbf{v}_{j}$, for DMF-based RSs, we plug in $\hat{r}^{\textrm{DMF}}_{ij} = f^{u}_{nn}(\mathbf{u}_{i})^{T} \cdot f^{v}_{nn}(\mathbf{v}_{j})$, etc. 

In practice, since the conditional unconfoundedness assumption in Eq. (\ref{eq:csi_assu}) is untestable, it is usually infeasible to calculate the exact value of $e_{ij}$ based on user/item covariates that satisfy Eq. (\ref{eq:csi_assu}).  Nevertheless, we can still calculate approximate propensity scores $\tilde{e}_{ij}$ and reweight the observed ratings by $1/\tilde{e}_{ij}$, but the unbiasedness of Eq. (\ref{eq:IPW}) after the reweighting cannot be guaranteed. Here we introduce two strategies for the approximate estimation. If user/item features $\mathbf{f}^{u}_{i}$ and $\mathbf{f}^{v}_{j}$ are available, $\tilde{e}_{ij}$ can be estimated with \textit{logistic regression} \cite{schnabel2016recommendations} as follows:
\begin{equation}
\label{eq:ps_lr}
    \tilde{e}^{\text{LR}}_{ij} = \text{Sigmoid}\Big(\Big(\sum_{k} w^{u}_{k}f^{u}_{ik}\Big) + \Big(\sum_{k} w^{v}_{k}f^{v}_{jk}\Big) + b_{i} + b_{j}\Big),
\end{equation}\noindent where $\text{Sigmoid}(x) = {(1+\mathrm{exp}(-x))}^{-1}$, $w^{u}_{k}$ and $w^{v}_{k}$ are the regression coefficients, and $b_{i}$, $b_{j}$ are the user and item-specific offsets, respectively. If user/item features $\mathbf{f}^{u}_{i}$ and $\mathbf{f}^{v}_{j}$ are not available, we can crudely approximate $e_{ij}$ based on the exposure data alone. For example, we can estimate $\tilde{e}_{ij}$ with \textit{Poisson factorization} \cite{liang2016causal} as:
\begin{equation}
\label{eq:ps_pf}
    \tilde{e}^{\text{PF}}_{ij} \approx 1 - \mathrm{exp}\left(-\boldsymbol{\pi}_{i}^{T} \cdot \boldsymbol{\gamma}_{j}\right),
\end{equation}
where $\boldsymbol{\pi_{i}}$ and $\boldsymbol{\gamma_{j}}$ are trainable user and item embeddings with Gamma prior, and they can be inferred from the exposure data as discussed in \cite{gopalan2015scalable}. Additional strategies to calculate the propensity scores can be found in \cite{ai2018unbiased,zhang2020large,wang2022unbiased,zhou2021contrastive}.

The advantage of IPW is that the unbiasedness of Eq. (\ref{eq:IPW}) for rating potential outcome estimation can be guaranteed if the propensity scores $e_{ij}$ are correctly estimated. However, the accuracy of the propensity score estimation models relies heavily on the domain knowledge and expertise of human experts, which is untestable by experiments. In addition, IPW suffers from a large variance and numerical instability issues, especially when the estimated propensity scores $e_{ij}$ are very small. Therefore, variance reduction techniques such as clipping and multi-task learning are usually applied to improve the stability of the training dynamics \cite{bottou2013counterfactual,saito2020unbiased, zhu2020unbiased}.

\vspace{0.2cm}
\noindent \textbf{Substitute Confounder Adjustment.} IPW-based RSs address exposure bias from the data's perspective: They reweight the biased observational dataset to create a pseudo randomized dataset that allows unbiased training of RSs. Confounder adjustment-based methods, in contrast, estimate confounders $C$ that cause the exposure bias and adjust their effects in the rating prediction model (A simple adjustment strategy is to control $C$ as extra covariates\footnote{Consider again the toy example in Fig. \ref{fig:RCM_IPW}. If we know exactly the user type and item type $\mathbf{c}$ for each user-item pair, the predictions can be unbiased even if the item exposures are non-randomized. }). For the adjustment to be unbiased, classical causal inference requires the conditional unconfoundedess assumption in Eq. (\ref{eq:csi_assu}) hold, i.e., no unobserved confounders \cite{imbens2015causal}, which is generally infeasible in practice. Fortunately, recent advances in multi-cause causal inference \cite{wang2019blessings} have shown that we can control substitute confounders estimated from item co-exposure data instead, where exposure bias can be mitigated with weaker assumptions. 

We use $\mathbf{a}_{i} = [a_{i1}, \cdots, a_{iJ}]$ to denote the exposure status of all $J$ items to user $i$, which can be viewed as a bundle treatment in clinical trials \cite{zou2020counterfactual}. Wang et al. \cite{wang2020causal} showed that if we can estimate user-specific latent variables $\boldsymbol{\pi}_{i}$, such that conditional on $\boldsymbol{\pi}_{i}$, the exposures of different items to the user are mutually independent,  controlling $\boldsymbol{\pi}_{i}$ can eliminate the influence of multi-cause confounders $\mathbf{c}_{i}^{m}$ (i.e., confounders that simultaneously affect the exposure of multiple items and ratings). A simple proof of the claim is that, if $\mathbf{c}_{i}^{m}$ can still influence $\mathbf{a}_{i}$ and $\mathbf{r}_{i}$ after conditioning on $\boldsymbol{\pi}_{i}$, since $\mathbf{c}_{i}^{m}$ is an unobserved common cause for the exposure of different items, $a_{ij}$ cannot be conditionally independent (see the discussion of the fork structure in section \ref{sec:atom}), which renders a contradiction. The rigorous proof can be found in \cite{wang2019blessings}. Wang et al. further assumed that $p(\mathbf{a}_{i}|\boldsymbol{\pi}_{i}) = \Pi_{j}p(a_{ij}|\boldsymbol{\pi}_{i}) = \Pi_{j}\text{Poission}(\boldsymbol{\pi}_{i}^{T} \cdot \boldsymbol{\gamma_{j}})$ and used the Poisson factorization to infer $\boldsymbol{\pi}_{i}$ and $\boldsymbol{\gamma_{j}}$. Afterward, exposure bias can be mitigated by controlling $\boldsymbol{\pi}_{i}$ as extra covariates in the RS model \cite{imbens2015causal}. For example, controlling $\boldsymbol{\pi}_{i}$ in MF-based RSs leads to the following adjustment:
\begin{equation}
\label{eq:scaj}
    r^{\text{adj}}_{ij}(a_{ij}=1) \sim \mathcal{N}\Big(\underbrace{\vphantom{\sum_{k}w_{k}\pi_{ik}} \mathbf{u}_{i}^{T} \cdot \mathbf{v}_{j}}_{\mathrm{user\ interests}} + \underbrace{\sum_{k}w_{k}\pi_{ik}}_{\mathrm{adj.\ for\\ \ expo.\ bias}}, \sigma_{ij}^{2}\Big).
\end{equation}
The property of propensity scores can be utilized to further simplify Eq. (\ref{eq:scaj}): If unconfoundedness in Eq. (\ref{eq:csi_assu}) holds for $C=\boldsymbol{\pi}_{i}$, it will also hold for $C = \tilde{e}_{ij} = p(a_{ij} | \boldsymbol{\pi}_{i})$ \cite{rosenbaum1983central}. Therefore, we can control the approximate propensity scores estimated by $\boldsymbol{\pi}_{i}$, i.e., $\tilde{e}_{ij} = \boldsymbol{\pi}_{i}^{T} \cdot \boldsymbol{\gamma_{j}}$, which leads to the simplified adjustment formula:
\begin{equation}
    r^{\text{adj}}_{ij}(a_{ij}=1) \sim \mathcal{N}\left(\mathbf{u}_{i}^{T} \cdot \mathbf{v}_{j} + w_{i} \cdot \tilde{e}_{ij}, \sigma_{ij}^{2}\right),
\end{equation}
where $w_{i}$ is a user-specific coefficient that captures the influence of $\tilde{e}_{ij}$ on ratings. 

Despite the success in addressing exposure bias with weaker assumptions, one limitation of the above method is that, since Poisson factorization is a shallow model, it may fail to capture the complex influences of multi-cause confounders on item co-exposures. To address this problem, recent works have introduced deep neural networks (DNNs) to infer the user-specific substitute confounders $\boldsymbol{\pi}_{i}$ from bundle treatment $\mathbf{a}_{i}$ \cite{zhu2022deep, ma2021multi}. These methods generally assume that $\mathbf{a}_{i}$ are generated from $\boldsymbol{\pi}_{i}$  via $p(\mathbf{a}_{i}|\boldsymbol{\pi}_{i})$ parameterized by a deep generative network $f^{\text{exp}}_{nn}$ as:
\begin{equation}
\label{eq:subs_deep}
    p(\mathbf{a}_{i}|\boldsymbol{\pi}_{i}) = \Pi_{j}\text{Bernoulli}(\text{Sigmoid}(f^{\text{exp}}_{nn}(\boldsymbol{\pi}_{i})_{j})),
\end{equation}
where the intractable posterior of $\boldsymbol{\pi}_{i}$ is then approximated with a Gaussian distribution parameterized by DNNs via the variational auto-encoding Bayes algorithm \cite{kingmaauto}, i.e., $
 q(\boldsymbol{\boldsymbol{\pi}_{i}} | \mathbf{a}_{i}) = \mathcal{N}(f^{\boldsymbol{\mu}}_{nn}(\mathbf{a}_{i}), \text{diag}(f^{\sigma^{2}}_{nn}(\mathbf{a}_{i})))$,
where $f^{\boldsymbol{\mu}}_{nn}$ and $f^{\boldsymbol{\sigma}^{2}}_{nn}$ are two DNNs that calculate the posterior mean and variance (before diagonalization) of $\boldsymbol{\pi}_{i}$. With deep generative models introduced to estimate the substitute confounders $\boldsymbol{\pi}_{i}$, non-linear influences of multi-cause confounders on item exposures can be adjusted in the RS models, where exposure bias can be further mitigated in recommendations.

The key advantage of substitute confounder estimation-based causal RSs is that controlling confounders in the potential outcome prediction model generally leads to lower variance than IPW-based methods \cite{wang2020causal}. However, these models need to estimate substitute confounders $\boldsymbol{\pi}_{i}$ from the item co-exposures and introduce extra parameters in the RS models to adjust their influences, which may incur extra bias if the confounders and the parameters are not correctly estimated. In addition, exposure bias due to single-cause confounders cannot be addressed by these methods. 

\subsubsection{Popularity Bias}
\label{sec:pop_bias}
 
Popularity bias can be viewed as a special kind of exposure bias where \textit{users are overly exposed to popular items} \cite{steck2011item, abdollahpouri2019unfairness}. Therefore, it can be addressed with techniques introduced in the previous section, especially the IPW-based methods \cite{zhu2021popularity}. The reason is that, if we define the popularity of an item as its exposure rate:
\begin{equation}
\label{eq:pop}
m_{j} = \frac{\sum_{i}a_{ij}}{\sum_{j}\sum_{i}a_{ij}},
\end{equation}
we can view $m_{j}$ as pseudo propensity scores and use IPW to reweight the observed ratings. Alternatively, we can also analyze and address popularity bias with the structural causal model (SCM), where the causal mechanism that generates the observed ratings under the influence of item popularity is deeply investigated.

\begin{figure*}[t]
\centering
\includegraphics[width=.68\textwidth,]{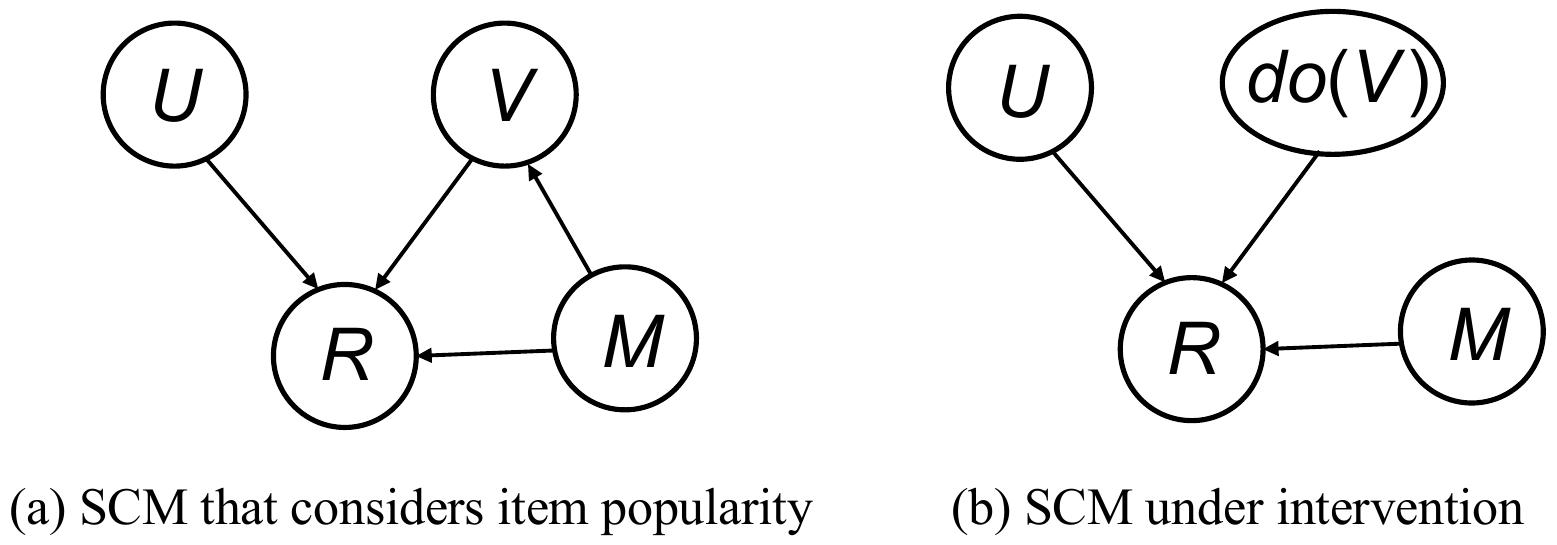}
      \caption{(a): SCM that explicitly models item popularity. (b): SCM under intervention $do(V)$.}
       \label{fig:pop_bias}
\end{figure*}

The discussion is mainly based on the popularity-bias deconfounding (PD) algorithm proposed in \cite{zhang2021causal}. PD assumes that the relations among user interests $\mathbf{u}_{i}$, item latent attributes $\mathbf{v}_{j}$, item popularity $m_{j}$, and observed ratings $r_{ij}$ can be represented by the causal graph illustrated in Fig. \ref{fig:pop_bias}, where item popularity can be clearly identified as a confounder that spuriously correlates the item attributes and the user ratings. PD aims to eliminate such spurious correlations with backdoor adjustment, such that the causal influences of $\mathbf{u}_{i}$ and $\mathbf{v}_{j}$ on $r_{ij}$ (which represents users’ interests on intrinsic item properties) can be properly identified. Recall that backdoor adjustment with SCM involves two stages: (1) During the training phase, the relevant structural equations in the causal graph are estimated from the collected dataset. 
(2) Afterward, we adjust the influence of confounders according to Eq. (\ref{eq:bkd_adj}) to remove the spurious correlations. Therefore, we need to estimate $p_{G}(r_{ij} | \mathbf{u}_{i}, \mathbf{v}_{j}, m_{j})$ with the observed ratings $r_{ij}$ and item popularty $m_{j}$ and infer the latent variables $ \mathbf{u}_{i}$ and $\mathbf{v}_{j}$. In PD, $p_{G}(r_{ij} | \mathbf{u}_{i}, \mathbf{v}_{j}, m_{j})$ is modeled as a variant of MF as follows:
\begin{equation}
\label{eq:popbias_now}
p_{G}(r_{ij} | \mathbf{u}_{i}, \mathbf{v}_{j}, m_{j}) \propto \underbrace{\text{Elu}(\mathbf{u}^{T}_{i} \cdot \mathbf{v}_{j}) \vphantom{m_{j}^{\lambda}}}_{\mathrm{user\ interests}}  \ \ \times \underbrace{m_{j}^{\lambda}}_{\mathrm{pop.\ bias}},
\end{equation}
where $\lambda$ is a hyper-parameter that denotes our belief toward the strength of influence of item popularity on ratings, and the function Elu (defined as $\text{Elu}(x) = e(x)$ if $x<0$ else $x+1$) makes the RHS of Eq. (\ref{eq:popbias_now}) a proper unnormalized probability density function. After $\mathbf{u}_{i}$, $\mathbf{v}_{j}$ are estimated from the datasets with Eq. (\ref{eq:popbias_now}), we conduct an intervention on the item node $V$ in the causal graph (see Eq. (\ref{eq:bkd_adj})), where the spurious correlation due to item popularity can be eliminated with backdoor adjustment:
\begin{equation}
\label{eq:bdj_popbias}
p(r_{ij} | do(\mathbf{u}_{i}, \mathbf{v}_{j})) \propto \mathbb{E}_{p(m_{j})}[\text{Elu}(\mathbf{u}^{T}_{i} \cdot \mathbf{v}_{j}) \times m_{j}^{\lambda}] =  \text{Elu}(\mathbf{u}^{T}_{i} \cdot \mathbf{v}_{j}) \times  \mathbb{E}_{p(m_{j})}[m_{j}^{\lambda}] .
\end{equation}
Since the second term $\mathbb{E}_{p(m_{j})}[m_{j}^{\lambda}]$ in Eq. (\ref{eq:bdj_popbias}) is a constant and $\text{Elu}$ is a monotonically increasing function, they have no influence on the ranking of the uninteracted items in the prediction phase. Therefore, we can drop them and use $\hat{r}_{ij} = \mathbf{u}^{T}_{i} \cdot \mathbf{v}_{j}$ as the unbiased rating predictor to generate future recommendations. 

Generally, the debiasing mechanism of PD is very intuitive and universal among backdoor adjustment-based causal RSs \cite{wang2021deconfounded,xu2021causal}: When fitting the RS model on the biased training set, we explicitly introduce the item popularity $m_{j}$ (i.e., the confounder) in Eq. (\ref{eq:popbias_now}) to explain away the spurious correlation between item attributes and the observed user ratings. Therefore, the user/item latent variables $\mathbf{u}_{i}$ and $\mathbf{v}_{j}$ used to generate future recommendations, i.e., $\hat{r}_{ij} = \mathbf{u}^{T}_{i} \cdot \mathbf{v}_{j}$, can focus exclusively on estimating users' true interests on intrinsic item properties.
\vspace{2mm}
\begin{svgraybox}\textbf{Is popularity bias always bad?} Recently, more researchers have begun to believe that popularity bias is not necessarily bad for RSs, because some items are popular because they \textit{per se} have better quality than other items or they catch the current trends of user interests, where more recommendations for these items can be well-justified \cite{zhao2022popularity, chen2022co}. Therefore, rather than setting the interventional distribution of item popularity to $p(m_{j})$, PD introduced above as well as some other methods \cite{zhang2021causal} further propose to make it correspond to item qualities or reflect the future popularity predictions. We will introduce these strategies in Section \ref{sec:causal_gen} regarding causal generalizations of RSs.
\end{svgraybox}

\subsubsection{Clickbait Bias}

Different from previous subsections that mainly focus on causal debiasing strategies for collaborative filtering-based RSs, this section discusses content-based recommendations. Specifically, we discuss the clickbait bias, which is defined as the bias of \textit{overly recommending items with attractive exposure features such as sensational titles but with low content qualities}. The discussion is mainly based on \cite{wang2021clicks}. We assume that item features $\mathbf{f}^{v}_{j}$ can be further decomposed into the item content feature $\mathbf{f}_{j}^{c}$ that captures item content information and the item exposure feature $\mathbf{f}_{j}^{b}$ whose main purpose is to attract users' attention. Taking micro-video as an example, item content feature $\mathbf{f}_{j}^{c}$ can be the audiovisual content of the video, whereas item exposure feature $\mathbf{f}_{j}^{b}$ can be its title, which is not obliged to describe its content faithfully. 

\begin{figure*}[t]
\centering
\includegraphics[width=.82\textwidth,]{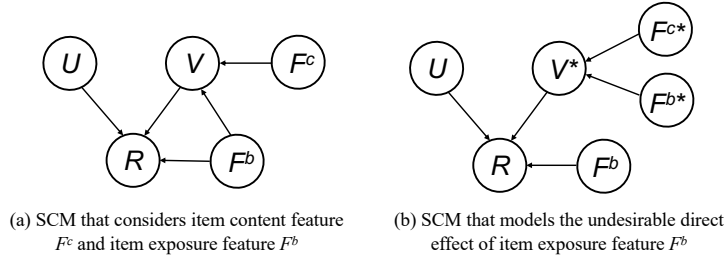}
      \caption{(a): The SCM that considers both the causal influences of item content feature $F^{c}$ and item exposure feature $F^{b}$ on item latent variable $V$. (b): The counterfactual SCM where $V^{*}$ is determined by baseline value $F^{b*}$ and $F^{c*}$ to model the undesirable direct effects of $F^{b}$.}
       \label{fig:clickbait-bias}
\end{figure*}

The relations among user interests $\mathbf{u}_{i}$, item exposure feature $\mathbf{f}_{j}^{b}$, item content feature $\mathbf{f}_{j}^{c}$, item fused features $\mathbf{v}_{j}$, and the observed ratings $r_{ij}$ are depicted in the causal graph in Fig. \ref{fig:clickbait-bias}-(a). We note that clickbait bias occurs when a user's recorded click on an item because she was cheated by the item exposure feature $\mathbf{f}_{j}^{b}$ before viewing the item content $\mathbf{f}_{j}^{c}$. Therefore, the bias can be defined as the \textbf{direct influence} of $\mathbf{f}_{j}^{b}$ on ratings $r_{ij}$ represented by the causal path $F^{b} \rightarrow R$. To eliminate the clickbait bias, we need to block the direct influence of $F^{b}$ on rating predictions, such that the item content quality can be comprehensively considered in recommendations.

As with SCM-based causal RSs, we first estimate structural equations of interest in the causal graph, i.e., $p_{G}(\mathbf{v}_{j}|\mathbf{f}^{b}_{j}, \mathbf{f}^{c}_{j})$ and $p_{G}(r_{ij}|\mathbf{u}_{i}, \mathbf{v}_{j}, \mathbf{f}^{b}_{j})$. Since distributions in \cite{wang2021clicks} are reasoned in a deterministic manner (i.e., Gaussian distributions with infinite precision), we keep the discussion consistent with them. Specifically, we use $\mathbf{v}_{j}(\mathbf{f}^{b}_{j}, \mathbf{f}^{c}_{j}) = f^{ff}(\mathbf{f}^{b}_{j}, \mathbf{f}^{c}_{j})$ to represent the structural equation $p_{G}(\mathbf{v}_{j}|\mathbf{f}^{b}_{j}, \mathbf{f}^{c}_{j})$, where $f^{ff}$ is the feature fusion function that aggregates $\mathbf{f}^{b}_{j}, \mathbf{f}^{c}_{j}$ into $\mathbf{v}_{j}$, and use $r_{ij}(\mathbf{u}_{i}, \mathbf{v}_{j}, \mathbf{f}^{b}_{j})$ to represent the structural equation $p_{G}(r_{ij}|\mathbf{u}_{i}, \mathbf{v}_{j}, \mathbf{f}^{b}_{j})$, respectively. To explicitly disentangle the influence of item exposure feature $\mathbf{f}^{b}_{j}$ and item latent variable $\mathbf{v}_{j}$ on the observed ratings, $r_{ij}(\mathbf{u}_{i}, \mathbf{v}_{j}, \mathbf{f}^{b}_{j})$ is assumed to factorize as follows:
\begin{equation}
\label{eq:bias}
    r_{ij}(\mathbf{u}_{i}, \mathbf{v}_{j}, \mathbf{f}^{b}_{j}) = \underbrace{f^{uv}_{nn}(\mathbf{u}_{i}, \mathbf{v}_{j}\vphantom{\text{Sigmoid}\left(f^{uf}_{nn}(\mathbf{u}_{i}, \mathbf{f}^{c}_{j})\right)})}_{\mathrm{user\ interests}} \cdot \underbrace{\text{Sigmoid}\left(f^{uf}_{nn}(\mathbf{u}_{i}, \mathbf{f}^{b}_{j})\right)}_{\mathrm{potential\ clickbait\ bias}},
\end{equation}
where the $\text{Sigmoid}$ function provides necessary non-linearity in the fusion process. Essentially, Eq. (\ref{eq:bias}) represents the causal mechanism that generates the observed ratings, which entangles both user interests in item content and clickbait bias. 

However, after learning the latent variables $\mathbf{u}_{i}, \mathbf{v}_{j}$ and functions $f^{uf}_{nn}, f^{uv}_{nn}$ via Eq. (\ref{eq:bias}), removing clickbait bias from the rating predictions is not as straightforward as the PD algorithm, because we should eliminate only the direct influence of item exposure feature $\mathbf{f}^{b}_{j}$ on ratings $r_{ij}$, while preserving its indirect influence mediated by item latent variable $\mathbf{v}_{j}$, such that all available item features can be comprehensively considered in recommendations. To achieve this purpose, we first calculate the natural direct effect (NDE) \cite{pearl2001direct} of item exposure feature $\mathbf{f}_{j}^{b}$ on ratings $r_{ij}$ as follows:
\begin{equation}
\label{eq:NDE}
    \text{NDE}(\mathbf{u}_{i}, \mathbf{v}^{*}_{j}, \mathbf{f}^{b}_{j}) =  r_{ij}(\mathbf{u}_{i}, \mathbf{v}^{*}_{j}, \mathbf{f}^{b}_{j}) - r_{ij}(\mathbf{u}_{i}, \mathbf{v}^{*}_{j}, \mathbf{f}^{b*}_{j}),
\end{equation}
where $\mathbf{v}^{*}_{j} = f^{ff}_{nn}(\mathbf{f}^{b*}_{j}, \mathbf{f}^{c*}_{j})$, and the baseline values $\mathbf{f}^{b*}_{j}$, $\mathbf{f}^{c*}_{j}$ are treated as if the corresponding features are missing from the item \cite{wang2021clicks}. Since the second term $r_{ij}(\mathbf{u}_{i}, \mathbf{v}^{*}_{j}, \mathbf{f}^{b*}_{j})$ in Eq. (\ref{eq:NDE}) denotes the user's rating to a “void” item and can be viewed as a constant, it will not affect the rank of the items. So we only adjust the first term of Eq. (\ref{eq:NDE}), which reasons with user $i$'s rating to item $j$ in a counterfactual world where item $j$ has only the exposure feature $\mathbf{f}^{b}_{j}$ but no content and fused features $\mathbf{f}^{c*}_{j}$ and $\mathbf{v}^{*}_{j}$, in Eq. (\ref{eq:bias})  (Fig. \ref{fig:clickbait-bias}-(b)). The adjustment leads to the following estimator,
\begin{equation}
\label{eq:adj_clickbait}
\hat{r}_{ij} =  r_{ij}(\mathbf{u}_{i}, \mathbf{v}_{j}, \mathbf{f}^{b}_{j}) - r_{ij}(\mathbf{u}_{i}, \mathbf{v}^{*}_{j}, \mathbf{f}^{b}_{j}) \triangleq  \underbrace{r_{ij}(\mathbf{u}_{i}, \mathbf{v}_{j}, \mathbf{f}^{b}_{j})}_{\text{user interests + clickbait}} - \ \ \underbrace{r_{ij}(\mathbf{u}_{i}, \mathbf{v}^{*}_{j}, \mathbf{f}^{b}_{j})}_{\text{clickbait bias}}.
\end{equation}
Eq. (\ref{eq:adj_clickbait}) removes the direct influence of $\mathbf{f}^{b}_{j}$ on rating predictions, such that item content quality can be comprehensively considered in future recommendations.

\subsubsection{Unfairness}

Recently, with the growing concern of algorithmic fairness, RSs are expected to show no discrimination against users from certain demographic groups \cite{li2021user, dong2022fairness, li2022fairness}. However, traditional RSs may capture the undesirable associations between users' sensitive information and their historical activities, which leads to potentially offensive recommendations to the users. Causal inference can help identify and address such unfair associations, where fairness can be promoted in future recommendations. This section focuses on the user-oriented fairness discussed in \cite{li2021towards}, which is defined as \textit{the bias where RS discriminately treats users with certain sensitive attributes}. 

\begin{figure*}[t]
\centering
\includegraphics[width=.54\textwidth,]{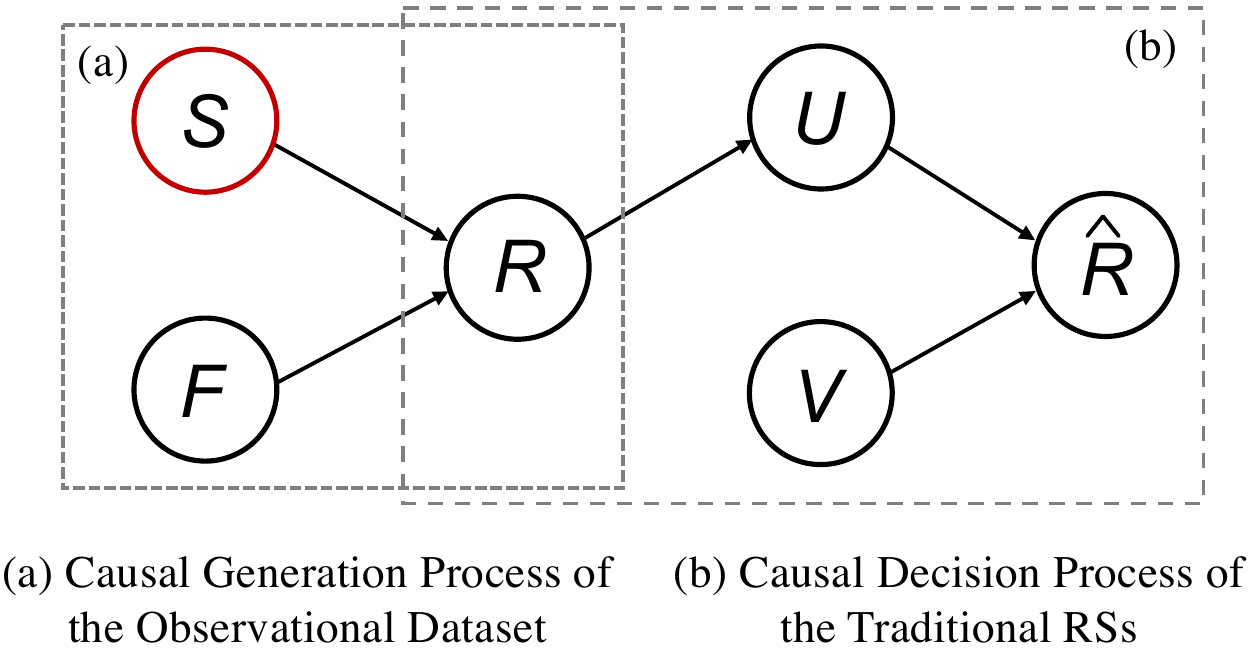}
      \caption{The SCM that reasons with the causal decision mechanism of traditional RSs. Observed user ratings $R$ can be causally driven by user features $F$, including sensitive features $S$, which can then unfairly influence the inference of user latent variables $U$ and new rating predictions $\hat{R}$.}
       \label{fig:fairness}
\end{figure*}

When considering the user-oriented fairness for RSs, a subset of user features $\mathbf{f}_{i}$, which we denote as $\mathbf{s}_{i}$, is assumed to contain the sensitive information of users, such as gender, race, and age. Features $\mathbf{s}_{i}$ are sensitive because recommendations that improperly rely on these features may be offensive to users, which degrade both their online experiences and their trust in the system. 
The causal graph that depicts the causal decision mechanism of most traditional RSs is illustrated in Fig. \ref{fig:fairness} \cite{li2021towards}. From Fig. \ref{fig:fairness} we can find that the user historical behaviors, i.e., the observed ratings $r_{ij}$, are causally driven by user features $\mathbf{f}_{i}$, including user sensitive features $\mathbf{s}_{i}$. Therefore, the user latent variables $\mathbf{u}_{i}$ inferred from $r_{ij}$ could capture sensitive user information in $\mathbf{s}_{i}$, which unfairly influences the rating predictions $\hat{r}_{ij}$ in the future. 

To address this problem, Li et al. \cite{li2021towards} proposed to disentangle the user sensitive features $\mathbf{s}_{i}$ from the user latent variable $\mathbf{u}_{i}$, such that the unfair influence of $\mathbf{s}_{i}$ on $\mathbf{u}_{i}$ represented by the causal chain $S \rightarrow R \rightarrow U$ can be maximally suppressed in the future recommendations. A common strategy to achieve the disentanglement is adversarial training \cite{goodfellow2020generative}, where we train a discriminator $f^{\text{cls}}_{nn}(\mathbf{u}_{i}) \rightarrow\mathbf{s}_{i}$ that predicts the sensitive features $\mathbf{s}_{i}$ from user latent variables $\mathbf{u}_{i}$ alongside the RS. While fitting the RS on the observe ratings $r_{ij}$, we constrain the inferred $\mathbf{u}_{i}$ to fool the discriminator $f^{\text{cls}}_{nn}$ by making wrong predictions about $\mathbf{s}_{i}$, which discourages $\mathbf{u}_{i}$ from capturing sensitive information in $r_{ij}$ due to its unfair correlations with $\mathbf{s}_{i}$. Here we take the MF-based RS as an example to show the details. We use $\mathcal{L}^{\text{Rec}}$ to denote the original training objective of the MF-based RS that maximizes the log-likelihood on observed ratings $r_{ij}$ and use $\mathcal{L}^{\text{cls}}$ to denote the loss function of the discriminator $f^{\text{cls}}_{nn}$. The adjusted training objective $\mathcal{L}^{\text{Fair}}$ with fairness constraint becomes the following:
\begin{equation}
\label{eq:adv}
    \mathcal{L}^{\text{Fair}} = \underbrace{\mathcal{L}^{\text{Rec}}(\mathbf{u}^{T}_{i} \cdot \mathbf{v}_{j}, r_{ij})}_{\mathrm{user\ interests}} - \lambda \cdot \underbrace{\mathcal{L}^{\text{cls}}(f^{\text{cls}}_{nn}\left(\mathbf{u}_{i}), \mathbf{s}_{i}\right)}_{\mathrm{fairness\ constraint}},
\end{equation}
where $\lambda$ is a hyper-parameter that balances the recommendation performance and the fairness objective. Generally, a higher $\lambda$ leads to better fairness, but it also restricts the capacity of the user latent variables $\mathbf{u}_{i}$, which could negatively impact the recommendation performance. Although here we use the MF-based RS as an example, it is straightforward to generalize Eq. (\ref{eq:adv}) to DMF or AE-based RS by replacing the $\mathbf{u}^{T}_{i} \cdot \mathbf{v}_{j}$ term with the corresponding rating estimators.

\subsection{Causal Explanation in Recommendations}

In previous sections, we have introduced causality to address various types of bias and spurious correlation issues for traditional RSs. In this section, we use causality to explain the user decision process. Specifically, we discuss an interesting question aiming to disentangle users' intent that causally explains their past behaviors, i.e., \textit{did a user purchase an item because she conformed to the current trend or because she really liked it}? The tricky part of this question is that: in reality, we only observe the effects, i.e., the purchases, which can be explained by both causes. 

\subsubsection{Disentangling Interest and Conformity with Causal Embedding}

\begin{figure*}[t]
\centering
\includegraphics[width=.7\textwidth,]{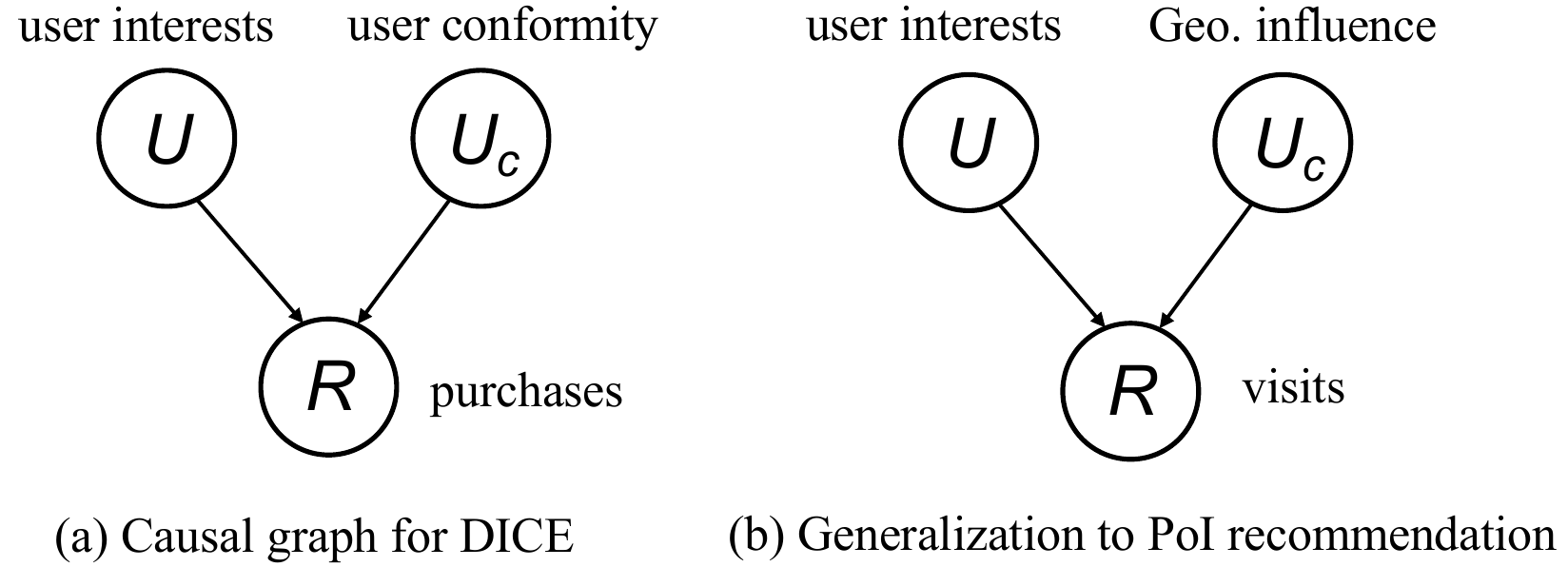}
      \caption{Causal Graphs for DICE (a) and its generalization to PoI recommendations (b).}
       \label{fig:dice}
\end{figure*}

The discussion is based on DICE proposed in \cite{zheng2021disentangling}. To simplify the discussion, we consider $r_{ij}$ as implicit feedback and define the set of user, positive item ($j:r_{ij}=1$), negative item ($k:r_{ik}=0$) triplets as $\mathcal{R}_{pn} = \{(i,j,k) | r_{ij}=1 \land r_{ik}=0 \}$. The popularity of each item $j$, i.e., $m_{j}$, which reflects the current trend, can be calculated with Eq. (\ref{eq:pop}). Observing that the causal relation between user interests $U$, user conformity $U_{c}$ and observed ratings $R$ can be represented as a V-structure in Fig. \ref{fig:dice}-(a), DICE exploits the \textit{colliding effect} to achieve the disentanglement, i.e., outcomes that cannot be explained by one cause are more likely caused by another (see discussions in Section \ref{sec:atom}). Therefore, although users' interests cannot be directly estimated from their ratings $r_{ij}$ due to entanglement, their conformity to the trend can be estimated by the popularity level of item $j$, and positive feedback not likely caused by conformity has a higher chance of reflecting users' true interests. 

In implementation, DICE assumes that the observed ratings $r_{ij}$ can be decomposed into the sum of a conformity part $r^{c}_{ij} = f^{c}(\mathbf{u}^{c}_{i}, \mathbf{v}^{c}_{j})$ and a user interests part $r^{i}_{ij} = f^{i}(\mathbf{u}^{i}_{i}, \mathbf{v}^{i}_{j})$, where $\mathbf{u}^{c,i}_{i}, \mathbf{v}^{c,i}_{j}$ are learnable user, item embeddings that reflect user $i$'s interests in (i.e., superscript $i$) and conformity to (i.e., superscript $c$) item $j$, respectively. According to the colliding effect of causal graphs, we can split the triplets in $\mathcal{R}_{pn}$ into two parts: In the first part $\mathcal{R}^{(1)}_{pn}$, positive item $a$ in the triplet has a higher popularity level than the negative item $b$, i.e., $m_{a} > m_{b}$. In this case, we can draw two general conclusions from this triplet: (1) Overall, the user prefers item $a$ over $b$; (2) She is more likely to conform to item $a$ than item $b$ due to $a$'s higher popularity. These conclusions lead to the two inequalities as follows:
\begin{equation}
\label{eq:1}
\forall (i,a,b) \in \mathcal{R}^{(1)}_{pn}\mathrm{, we\ have}
\begin{cases}
\ r_{i a}^c>r_{i b}^c \ (\mathrm{conformity}) \\
\ r_{i a}^i+r_{i a}^c>r_{i b}^i+r_{i b}^c \ (\mathrm{overall\ preference}),
\end{cases}    
\end{equation}
where the dependency of $r^{c,i}_{i\{a, b\}}$ on latent variables $\mathbf{u}^{c,i}_{i}, \mathbf{v}^{c,i}_{\{a, b\}}$ are omitted for simplicity. The second part, i.e., $\mathcal{R}^{(2)}_{pn}$, is the \textbf{key} to achieving disentanglement, because for every triplet $(i,c,d)$ in $\mathcal{R}^{(2)}_{pn}$, the negative item $d$ is more popular than the positive item $c$. In this case, user $i$ \textit{could have simply conformed to the trend} and chosen item $d$ to consume, but instead, she actively chose the less popular item $c$. Therefore, we can draw one more specific conclusion that leads to the disentanglement between user interests and conformity: The choice of item $c$ over $d$ is more likely due to user interests. Therefore, we can form three inequalities as:
\begin{equation}
\label{eq:2}
\forall (i,c,d) \in \mathcal{R}^{(2)}_{pn}\mathrm{, we\ have}
\begin{cases}
\ r_{i c}^i>r_{i d}^i\ (\mathrm{interests}), r_{i c}^c<r_{i d}^c\ (\mathrm{conformity}), \\
\ r_{i c}^i+r_{i c}^c> r_{i d}^i+r_{i d}^c\ (\mathrm{overall\ preference}).
\end{cases} 
\end{equation}
The inequalities in Eqs. (\ref{eq:1}) and (\ref{eq:2}) can be solved by ranking-based loss in RSs, such as Bayesian personalized ranking (BPR) \cite{rendle2009bpr}, where the disentangled embeddings $\mathbf{u}^{c,i}_{i}, \mathbf{v}^{c,i}_{j}$ and the match functions $f^{c,i}(\cdot, \cdot)$ can be learned from $\mathcal{R}^{(1)}_{pn}$ and $\mathcal{R}^{(2)}_{pn}$. Finally, we form a rating predictor $\hat{r}_{ij} = f^{i}(\mathbf{u}^{i}_{i}, \mathbf{v}^{i}_{j}) + f^{c}(\mathbf{u}^{c}_{i}, \mathbf{v}^{c}_{j})$ for future recommendations. 

\subsubsection{Generalizations of DICE}

DICE disentangles the user intent and promotes the explainability of RSs from the data's perspective: It partitions the triplets $(i,j,k)$ in $\mathcal{R}_{pn}$ into two disjoint subsets $\mathcal{R}^{(1)}_{pn}$ and $\mathcal{R}^{(2)}_{pn}$ based on the relative popularity of the positive and negative items, and shows that the triplets in $\mathcal{R}^{(2)}_{pn}$ are informative to distinguish the user interests from their conformity to the popularity trend. The basic idea of DICE is generalizable to promote explainability for other types of recommendation tasks, if we can find \textbf{alternative causal explanations} to challenge the assumption that the observed positive feedback in these tasks can be attributed solely to user interests. 

For example, in point-of-interests (PoI) recommendations, the target items are specific point locations that users may find useful or interesting to visit, such as restaurants, grocery stores, and malls \cite{ye2011exploiting}. In this task, the location of a PoI is an important alternative explanation for users' visits to the PoI other than user interests, because nearby POIs are more convenient to visit than the remote ones \cite{wang2018exploiting}. Therefore, to disentangle user interests from potential geographical factors that could causally influence users' choices, we can take a similar strategy as DICE and partition the user historical visit records according to the distance of positive and negative PoIs to users. Then, the disentangled user interest embeddings can be estimated based on the partitioned dataset with the same ranking-based approach.

\subsubsection{Other Works on Explainable RSs}

Explanable recommendation is a broad topic \cite{zhang2020explainable}, where disentangling user's intent based on data partitioning is a small part. There are also plenty of works that focus on improving the explainability of RSs from the model's side, where specific disentanglement modules, such as prototype learning \cite{ma2019learning}, context modeling \cite{wang2022causal}, and aspect modeling \cite {tan2021counterfactual}, are designed and integrated with traditional RS models to further enhance their transparency and explainability. We refer interested readers to the corresponding papers as well as \cite{xu2021learning,sheth2022causal} for further investigation.

\subsection{Causal Generalization of Recommendations}
\label{sec:causal_gen}

After estimating the causal relations from potentially biased and entangled observational datasets, the generalization ability of RSs can be substantially enhanced, because even if the context (or environment) in which we make recommendations changes (e.g., item popularity, user conformity, etc.), we can still basing the recommendations on causal relations that are \textit{stable and invariant}, while discarding or correcting other undesirable correlations that are transient and susceptible to change \cite{zheng2021disentangling,arjovsky2019invariant}. In this section, we use the PD algorithm for popularity bias
 and the DICE algorithm for causal explainability as two examples to show how the generalization of RSs can be improved with causal intervention and disentanglement.
 
\subsubsection{Generalization Based on Intervention}
First, we take the PD algorithm as an example to show how causal intervention can improve the generalization of RSs within a dynamic environment. In RS, it is generally assumed that user interests can remain unchanged for a certain period of time, i.e., the causal structure $U \rightarrow R \leftarrow V$ in Fig. (\ref{fig:pop_bias}) represents the stable user interests on intrinsic item properties. However, the popularity of different items, i.e., the context in which we make recommendations, can shift rapidly during the same period \cite{chen2022co}. Recall that PD disentangles the causal influences of user interests and item popularity on ratings via the product of two terms, i.e., $\text{Elu}(\mathbf{u}^{T}_{i} \cdot \mathbf{v}_{j})$ and $m_{j}^{\lambda}$, as Eq. (\ref{eq:popbias_now}). Suppose $m_{j}$ represents the current popularity level of item $j$. If we predict that the popularity of item $j$ will change to $m^{\prime}_{j}$ in the future \cite{xie2020multimodal}, we can conduct an intervention that sets $M$ to the predicted value $m_{j}^{\prime}$ in the structural equation $p_{G}(R|U,V,M)$ and predict future ratings $r^{\prime}_{ij}$ via the following formula:
\begin{equation}
p_{G}(r^{\prime}_{ij} | \mathbf{u}_{i}, \mathbf{v}_{j}, do(m^{\prime}_{j})) \propto \underbrace{\vphantom{(m^{\prime}_{j})^{\lambda}}\text{Elu}(\mathbf{u}^{T}_{i} \cdot \mathbf{v}_{j})}_{\mathrm{stable\ user\ interests}} \times  \underbrace{(m^{\prime}_{j})^{\lambda}}_{\mathrm{future \ popularity}},
\end{equation}
where the user, item latent variables $\mathbf{u}_{i}$ and $\mathbf{v}_{j}$ learned from the current time step remain unaltered. With the influence of future changes in item popularity on ratings considered in the predictions, service providers can make strategic decisions to allocate resources for items with different popularity potentials. In contrast, traditional RSs could mistakenly capture the influence of the current popularity level of items on ratings as user interests. Therefore, they will not generalize well when the item popularity $m_{j}$ changes to a different level $m^{\prime}_{j}$ due to time evolution.

\subsubsection{Generalization Based on Disentanglement}
In addition, causal disentanglement can promote the generalization of RSs by identifying and basing recommendations on causes that are more robust to potential changes in the environments \cite{suter2019robustly,yang2021causalvae}. For example, if users' conformity and interest are disentangled based on their historical behaviors, if a user's conformity reduces to a low level due to certain reasons, since user interests are assumed to be stable within a certain period of time, we can still use the learned user/item interest variables $\mathbf{u}^{i}_{i}$, $\mathbf{v}^{i}_{j}$ to make recommendations based on their interests, where the previously estimated unreliable user conformity information can be discarded or down-weighted. In contrast, for traditional RSs, different factors that causally influence their historical behaviors are entangled as a single user latent variable $\mathbf{u}_{i}$. Therefore, even if some less stable causes of user behaviors are known to change (e.g., in the PoI RS introduced above, a user could move to a new place where the convenience levels of different PoIs change for the user), these models will still utilize the outdated causes to make recommendations, which could fail to generalize to the new environment.

\section{Evaluation Strategies for Causal RSs}
\label{sec:datasets}

In the previous sections, we have discussed various causal inference techniques that are promising to address multiple types of biases, entanglement, and generalization problems in traditional RSs. However, without a well-designed model evaluation strategy, it is difficult to tell whether the proposed causal RS model is indeed effective, nor can we guarantee that the model will perform reliably after deploying in a real-world environment. The evaluation of causal models is particularly difficult, because the groundtruths, i.e., the causal effects of interest, are usually infeasible. Despite the challenges, there are several strategies that can reliably evaluate causal RSs with biased real-world data, and we will thoroughly discuss them in this section. In addition, we also compile the available real-world datasets that conduct randomized experiments to eliminate exposure bias to facilitate future causal RS research.

\subsection{Evaluation Strategies for Traditional RSs}

The assessment of traditional RSs generally follows three steps: First, the observed ratings $r_{ij}$ in the rating matrix $\mathbf{R}$ are split into the non-overlapping training set $\mathbf{R}_{tr}$ and test set $\mathbf{R}_{te}$, usually by randomly holding out a certain percentage of the observed ratings from each user. Then, the proposed RS is trained on ratings in $\mathbf{R}_{tr}$ to learn the latent variables and the associated functional models (see Section \ref{sec:rs_basics}). Finally, the trained RS predicts the missing ratings in $\mathbf{R}_{tr}$ for each user, where the results are compared with the held-out ratings in $\mathbf{R}_{te}$ to evaluate the model performance. The quality of rating predictions can be measured by accuracy-based metrics such as mean squared error (MSE) and mean absolute error (MAE), and ranking-based metrics such as recall, precision, normalized discounted cumulative gain (NDCG), etc. More information on these evaluation metrics can be found in \cite{shani2011evaluating}.

\subsection{Challenges for the Evaluation of Causal RSs}

The above evaluation strategy, however, is not directly applicable to causal RSs, because ratings in $\mathbf{R}_{te}$ may have the same spurious correlation and bias as ratings in $\mathbf{R}_{tr}$, which makes the evaluation on $\mathbf{R}_{te}$ a biased measure of the true model performance. Therefore, to unbiasedly evaluate the effectiveness of causal RSs, it is ideal that we have a biased/entangled training set $\mathbf{R}^{b}_{tr}$ to learn the model, and an unbiased/disentangled test set $\mathbf{R}^{ub}_{te}$ for model evaluation, such that the effectiveness of the causal debiasing/disentangling algorithm can be directly verified from experiments. However, such unbiased/disentangled test set $\mathbf{R}^{ub}_{te}$ can be difficult to acquire and expansive to establish. Therefore, we first introduce common data simulation strategies for causal RS evaluation. We then discuss how real-world datasets can be directly utilized to further promote the credibility of causal RS research. 

\subsection{Evaluation Based on Simulated Datasets}

A good dataset simulation strategy to evaluate causal RSs should have the following properties: (1) The generation mechanisms of the bias and entanglement to be studied are clearly identified, credibly designed, and can be adjusted in a flexible manner; (2) The available real-world information is utilized as much as possible. 

\subsubsection{Simulation Based on Generative Models}

One promising dataset simulation strategy that satisfies the above criteria is to use deep generative models. Here we take exposure bias as an example to demonstrate how it can be simulated from real-world datasets \cite{zhu2022deep}. The simulation is composed of two phases. In the training phase, two variational auto-encoders (VAEs) \cite{liang2018variational,kingmaauto} are trained on the exposure and rating data in a real-world dataset (e.g., the MovieLens dataset \cite{harper2015movielens}), which results in two decoder networks $f^{a}_{nn}$ and $f^{r}_{nn}$ that generate item exposures $\mathbf{a}_{i} \in \{0, 1\}^{J}$ and user ratings $\mathbf{r}_{i}  \in \mathbb{R}^{J}$ from $K$-dimensional Gaussian user latent variables $\mathbf{u}^{a}_{i} \sim \mathcal{N}(\mathbf{0}, \mathbf{I}_{K})$ and $\mathbf{u}^{r}_{i} \sim \mathcal{N}(\mathbf{0}, \mathbf{I}_{K})$, respectively. The decoders capture the generative distributions of item exposures and user ratings based on the data of real users, where the available real-world information is effectively utilized. In the generation phase, for each hypothetical user $i'$, we draw a confounder $\mathbf{c}_{i'} \sim \mathcal{N}(\mathbf{0}, \mathbf{I}_{K})$ that simultaneously affects $\mathbf{u}^{a}_{i'}$ and $\mathbf{u}^{r}_{i'}$. Then, to simulate the exposure bias, we set $\mathbf{u}^{a}_{i'} = \mathbf{c}_{i'}$ and $\mathbf{u}^{r}_{i'} = \lambda \cdot \mathbf{c}_{i'} + (1-\lambda) \boldsymbol{\epsilon}_{i'}$ and use $f^{a}_{nn}$, $f^{r}_{nn}$ to generate the simulated item exposures $\mathbf{a}_{i'}$ and ratings $\mathbf{r}_{i'}$, where $\boldsymbol{\epsilon}_{i'} \sim \mathcal{N}(\mathbf{0}, \mathbf{I}_{K})$ and hyper-parameter $\lambda$ controls the strength of the confounding bias. Finally, we mask $\mathbf{r}_{i'}$ with $\mathbf{a}_{i'}$ to form the biased training set $\mathbf{R}^{b}_{tr}$, and keep the generated ratings $\mathbf{r}_{i'}$ unmasked in the test set $\mathbf{R}^{ub}_{te}$ for an unbiased evaluation of model performance. 

The advantage of dataset simulation strategies based on generative models is that the true causal mechanisms of interest, such as the rating potential outcomes, are available in the evaluation stage, which is generally impossible for real-world datasets. Therefore, the effectiveness of causal RSs can be easily verified based on the simulated groundtruths. In addition, the simulations are flexible as the strength of biases and entanglements can be set into different levels (e.g., $\lambda$ in the example), where the sensitivity and robustness of causal RSs can be thoroughly investigated.

\subsubsection{Test Set Intervention}

Another reliable dataset simulation strategy is test set intervention, where an intervened test set is created from the original test set, such that it has a different bias/entanglement distribution from the training set \cite{zheng2021disentangling,liang2016causal,zhu2022deeptag}. For example, to study the popularity bias, we can first select observed ratings from $\mathbf{R}$ such that 90\% of the interacted items are popular and 10\% are unpopular to form the training set $\mathbf{R}_{tr}$ \cite{yi2022debiased}. We then select from the remaining ratings, i.e., the original test set $\mathbf{R}_{te}$, a subset with a different ratio of popular and unpopular items (e.g., 10\% popular and 90\% unpopular) to form the intervened test set $\mathbf{R}^{int}_{te}$. If the causal RSs trained on $\mathbf{R}_{tr}$ can still perform well on the intervened test set $\mathbf{R}^{int}_{te}$,  the model's invariance to the popularity bias can be supported. A similar test set intervention strategies can be used to evaluate the disentanglement of user interests and conformity for DICE \cite{zheng2021disentangling}. 

The advantage of the test set intervention-based causal RS evaluation strategy is that extra assumptions that cannot be justified by real-world information are minimally introduced, because the intervention is usually achieved by selecting samples from the original test set to change the data distribution, which does not introduce extra assumptions of the generative mechanisms or hypothetical users, items, and ratings. From this perspective, the evaluation results based on test set intervention may be more credible compared with the generative model-based strategies. 

\subsection{Evaluation Based on Real-world Datasets}

\subsubsection{Randomized Experiments}
For the study of exposure bias, it is feasible to establish-bias free real-world datasets, where ratings for either every item or randomly selected items are collected from a subset of users. This can be extremely expansive and user-unfriendly, but recent years have witnessed a growing interest in causal RS research from the industry, where more such randomized datasets are established and released to facilitate causal RS research. The  available real-world datasets are compiled as follows:
\begin{itemize}[parsep=6pt]
    \item \textbf{Coat datasets}\footnote{\url{https://www.cs.cornell.edu/~schnabts/mnar/}
} \cite{schnabel2016recommendations} (2016). The Coat dataset is a small-scale dataset crowdsourced from the Amazon Mechanical Turkers platform with 300 users and 290 items. Specifically, each Turker is first asked to self-select 24 coats to rate, where the ratings form the biased training set $\mathbf{R}^{b}_{tr}$. Then each Turker is asked to rate 16 random coats, and these ratings form the unbiased test set $\mathbf{R}^{ub}_{te}$.
    \item \textbf{Yahoo! R3 dataset}\footnote{\url{https://webscope.sandbox.yahoo.com/catalog.php?datatype=r&did=3}} \cite{marlin2007collaborative,marlin2009collaborative} (2009). The Yahoo! R3 dataset is collected from the Yahoo! Music platform. The biased training set $\mathbf{R}^{b}_{tr}$ is composed of 300,000 self-supplied ratings from 15,400 users to 1,000 items. In addition, a subset of 5,400 users is presented with ten randomly selected items to rate, and the ratings are used to create the unbiased test set $\mathbf{R}^{ub}_{te}$.
    \item \textbf{KuaiRec dataset}\footnote{\url{https://github.com/chongminggao/KuaiRec}} \cite{gao2022kuairec} (2022). The KuaiRec dataset is established based on a popular micro-video sharing platform, KuaiShou, in China (known as Kwai internationally). The dataset records self-supplied ratings from 7,176 users to 10,728 items as the biased training set $\mathbf{R}^{b}_{tr}$. The unbiased test set $\mathbf{R}^{ub}_{te}$ is composed of a subset of 1,411 users and 3,327 items, where the ratings between these users and items are almost fully observed (with 99.6\% density).
\end{itemize}
The statistics of the datasets are summarized in Table \ref{tab:datasets} for reference. There are also randomized datasets for some related topics such as click-through rate prediction \cite{zhou2018deep}, i.e., Criteo Ads datasets\footnote{\url{http://cail.criteo.com/criteo- uplift- prediction- dataset/}} \cite{diemert2018large}, and bandit-based RS \cite{bouneffouf2012contextual}, i.e., Open Bandit dataset\footnote{\url{https://research.zozo.com/data.html}} \cite{saito2020large}, where the sources are also provided in case the readers are interested.
\begin{table}[]
    \centering
    \begin{tabular}{lccccc}
    \hline
     Dataset    &  \# Users & \# Items & Item Type & Training Sets & Test Sets\\
    \hline
     Coat    & 300 & 290 & Coat & 24 i/u (self-supplied)  & 16 i/u (random) \\
     Yahoo! R3 & 15,400 & 1,000 & Music & 300,000 r (self-supplied) & 10 i/u (random) for 5,400 u\\
     KuaiRec & 7,176 & 10,728 & Video & 16.3\% r (self-supplied) & 99.6\% r for 1,411 u and 3,327 i\\
     \hline
    \end{tabular}
    \caption{Characteristics of the currently available real-world causal recommendation datasets, where the test sets are devoid of exposure bias either due to randomized item exposures or fully observed ratings. In the table, terms like 24 i/u mean that every user rates 24 items, the term 300,000 r denotes the number of observed ratings, and terms like 16.3\% r represent the density of interactions.}
    \label{tab:datasets}
\end{table}

From Table \ref{tab:datasets} we can find that, the Coat dataset is small in scale. While for the Yahoo! R3 dataset, the training set is comparatively large (15,400 users and 1,000 items), the randomized experiment conducted to establish the unbiased test set is small-scale in comparison (16 and 10 randomly exposed items per user, respectively). Therefore, although these ratings are unbiased due to randomization, they may not capture well-rounded user interests and therefore induce a high evaluation variance. For the recently released KuaiRec datasets, large-scale experiments are conducted on users to establish the bias-free test set, where the 1,411 users' ratings for 3,327 items are almost fully collected. Therefore, it may be a promising new benchmark that allows the evaluation of more complex causal RS models with a lower variance.

\subsubsection{Qualitative Evaluation and Case Study}
For other types of biases in RSs that cannot be attributed to non-randomized item exposures (e.g., clickbait bias and unfairness), the establishment of bias-free test sets is more challenging. For example, when studying the clickbait bias, it is difficult to determine whether a user clicked an item due to interests or clickbait. Similarly, when examining the user-oriented fairness of RSs, we cannot know if the generated items are offensive to the users. Under such circumstances, we can still conduct case studies for qualitative model evaluations, where we manually select some representative samples from the original test set and observe whether the trained causal RS model would respond as expected to these samples \cite{wang2022causalRep}.

Consider the evaluation of the robustness of a causal RS to clickbait bias. We can select some representative items with low-quality content but highly-attractive exposure features and other items with high-quality content but normal exposure features from the original test set. Then, we obtain rating predictions for items from these two groups and draw comparisons. If the studied causal RS indeed ranks items in the second group higher than those in the first group, we can likely conclude that the model is robust to clickbait bias because the quality of the item content, not its exposure features, is prioritized in recommendations. In addition, to evaluate the user-oriented fairness of a causal RS, we can analyze the generated recommendation for users from certain demographic groups. If the recommended items tend to capture the social stereotypes that are negatively associated with user sensitive features, we can conclude that the model is still discriminatory against users.

\section{Future Directions}
\label{sec:future}

Despite the recent achievements in marrying causal inference with traditional RSs to address their various limitations of correlational reasoning on observational user data, causal RS research is still in its emerging stage. Several promising directions could be pursued to further advance this field. In this section, we identify four interesting and important open problems worthy of exploration in the future.

First, the assumptions required by existing causal RSs could be too strong, which may not hold in reality. For example, most RCM-based causal RSs rely on SUTVA to exclude the interference of item exposures for different users. However, if users are connected by a social network, they may interact closely with each other or be heavily affected by the influencers in the network \cite{ma2022learning2}. Consequently, SUTVA can be violated because the recommendations made to one user may causally affect the ratings of others (i.e., the spill-over effects \cite{li2022causal, ma2022learning}). In addition, the positivity assumptions may also be violated if some users never click certain types of items (i.e., non-compliance and defiers \cite{imbens2015causal}). Therefore, it is crucial to further weaken the assumptions of causal RSs to make them more practical for real-world applications.

In addition, there currently lacks a universal causal model for RSs that can be applied for different causal reasoning purposes. Most SCM-based causal RSs are designed to address one specific type of bias or entanglement problem, where other issues are tacitly assumed to be absent and omitted from the causal graph. Moreover, even for causal RSs that address the same problem, several varieties of causal graphs that include different sets of variables and relationships can be assumed, which leads to inconsistency between different works. Therefore, it would be promising and beneficial to have a generic and widely-accepted causal model that is able to comprehensively address multiple types of causal problems in recommendations.

Furthermore, certain types of biases in RSs are double-blade swords, where the positive side is seldom investigated. Consider the item exposure bias discussed in Section \ref{sec:exp_bias}. We should note that some items are more likely to be exposed because they have higher quality than other items. Therefore, the higher exposure rate of these items can be well justified and may be utilized to further enhance the recommendation performance. In addition, recent research also found that confounders that spuriously correlate item exposures and user ratings may also help explain the co-occurrence patterns of different items \cite{zhu2022deep}. Therefore, how to properly identify and utilize the positive side of biases while maximally suppressing their negative effects is of great importance and deserves more in-depth investigations in the future.

Finally, although recent years have witnessed the establishment and release of more real-world datasets for causal RS research from the industry, many causal RS models still rely heavily on simulated datasets for evaluation. The simulation can lead to the over-simplification of the problem and is often designed to correspond exactly with the debiasing/disentanglement mechanism of the proposed model. Therefore, the effectiveness of these methods in more complicated real-world scenarios is still uncertain due to the lack of model deployment and online tests. As such, to more convincingly demonstrate the practical utility of causal RSs, more collaborations with the industry are highly expected.

\section{Summary}
\label{sec:conclusions}
In this survey, we provide a comprehensive overview of recent advances in causal inference for RSs. We start by pointing out issues of traditional RSs that rely on correlations in observed user behaviors and user/item features. We then introduce two mainstream causal inference frameworks, i.e., Rubin's RCM and Pearl's SCM, which provide deeper insights into these issues and the foundation for moving traditional RSs to the upper rungs of the Ladder of Causality. Specifically, we thoroughly discuss several state-of-the-art causal RS models that lead to enhanced robustness to various biases and improved explainability. In addition, since causal RSs can base recommendations on causal relationships that are stable and invariant, we also demonstrate that their generalization abilities can be significantly improved. Finally, we introduce evaluation strategies for causal RSs, with an emphasis on how to reliably estimate the model performance based on biased real-world data. We further compile real-world datasets where expensive randomized experiments are conducted on users, which reflects growing attention to causal RSs from the industry.

Overall, causal RS is still a relatively new and under-explored research topic. More efforts are urgently demanded to systematize the existing works and conduct deeper investigations for further improvements. Accordingly, we point out four interesting and practically important open problems in causal RSs. We hope that this survey can help readers gain a comprehensive understanding of the main idea of applying causality in RSs and encourage further progress in this promising area. 

\vspace{0.2cm}

\noindent \textbf{Acknowledgements.} This work is supported by the National Science Foundation under grants IIS-2006844, IIS-2144209, IIS-2223769, CNS-2154962, and BCS-2228534, the JP Morgan Chase Faculty Research Award, and the Cisco Faculty Research Award.

\bibliographystyle{unsrt}
\bibliography{CI4RecSys}
\end{document}